\documentclass[aps,groupedaddress,preprint,nofootinbib]{revtex4-1}

\usepackage{amssymb}
\usepackage{amsbsy}
\usepackage{rotating}
\usepackage{amsmath}
\usepackage{subfigure}
\usepackage{rotating}
\usepackage[latin1]{inputenc}
\usepackage{amsthm,mathrsfs,amsopn}

\usepackage{graphicx}

\begin{document}

\title{\large{Topology-driven instabilities: the theory of pattern formation on
directed networks}}

\author{Malbor Asllani$^{1}$, Joseph D. Challenger$^{2}$,
Francesco Saverio Pavone$^{2,3,4}$, Leonardo Sacconi$^{3,4}$, Duccio Fanelli$^{2}$}
\affiliation{
1. Dipartimento di Scienza e Alta Tecnologia, University of Insubria, via Valleggio 11, 22100 Como, Italy \\
2. Dipartimento di Fisica e Astronomia, University of Florence and INFN, Via Sansone 1, 50019 Sesto Fiorentino, Florence, Italy \\
3. European Laboratory for Non-linear Spectroscopy, 50019 Sesto Fiorentino, Florence, Italy \\
4. National Institute of Optics, National Research Council, 50125 Florence, Italy 
} 
\begin{abstract}
The theory of pattern formation in reaction-diffusion systems is extended to the case of a directed network. Due to the structure of the network Laplacian of the scrutinised system, the dispersion relation has both real and imaginary parts, at variance with the conventional case for a symmetric network. It is found that the homogeneous fixed point can become unstable due to the topology of the network, resulting in a new class of instabilities which cannot be induced on undirected graphs. Results from a linear stability analysis allow the instability region to be analytically traced. Numerical simulations show that the instability can lead to travelling waves, or quasi-stationary patterns, depending on the characteristics of the underlying graph. 
The results presented here could impact on the diverse range of disciplines where directed networks are found, such as neuroscience, computer networks and traffic systems.
\end{abstract}

\maketitle

\vspace{0.8cm}

Self-organised structures can spontaneously emerge from a complex sea of microscopically interacting constituents. 
This is a widespread observation in nature, now accepted as a paradigmatic concept in science, with many interdisciplinary 
applications ranging from biology to physics. Hence, pinpointing key processes that yield macroscopically ordered patterns is a fascinating field of investigation, at the forefront of many exciting developments. 

For example, the Turing instability \cite{turing} constitutes a universal paradigm for the spontaneous generation of spatially organised patterns. It formally applies to a wide category of phenomena, which can be modelled via reaction-diffusion schemes \cite{mimura,maron,baurmann,riet,meinhardt,harris,maini,bhat,miura}. These are mathematical models that describe the coupled evolution of spatially distributed species, driven by microscopic reactions and freely diffusing in the embedding medium. Diffusion can potentially seed the instability by perturbing the homogeneous mean-field state, through an activator-inhibitor mechanism, resulting in the emergence of a patched, spatially inhomogeneous, density distribution. The most astonishing examples are perhaps encountered in the context of morphogenesis, the branch of embryology devoted to investigating the development of patterns and forms in biology \cite{murray}. Moreover, spatial patterns of cortical activation observed with EEG, MEG and fMRI have been argued to originate from the spontaneous self-organisation of interacting populations of excitatory and inhibitory neurons \cite{neuron0}. In this case, travelling waves can also develop following a symmetry breaking instability of a homogeneous fixed point \cite{zhab}. The models are usually defined on a regular lattice, which is either continuous or discrete, and the emphasis is placed on the specific form of the nonlinear interactions which are ultimately responsible for the occurrence of the instability. 

For a large class of problems, including applications to neuroscience, the system under scrutiny is instead defined on a complex network, rather than on a regular, spatially extended lattice. In a seminal paper \cite{nakao}, Nakao and Mikhailov investigated the effects of the embedding graph structure on the emergence of Turing patterns in nonlinear diffusive systems, paving the way for novel discoveries in an area of widespread interest. The conditions for the deterministic instability are derived following a linear stability analysis, which requires expanding the perturbation on the complete basis formed by the eigenvectors of the discrete Laplacian. Travelling waves can also develop via a completely analogous mechanism \cite{asllani}. To date, symmetric (undirected) networks have been considered in the literature. Again, the instability is driven by the non-trivial interplay between the nonlinearities, which are accounted for in the reaction parts, and the diffusion terms. The topology of the space defines the relevant directions for the spreading of the perturbation, but cannot impact on the onset of the instability. Consider for instance a reaction-diffusion system defined on a regular lattice and assume that it cannot experience a Turing-like instability: the system cannot be made unstable when placed on top of a symmetric network. In other words, the structure of the underlying graph cannot destabilise an otherwise stable scheme: the inherent ability of the model to self-organise in space and time is determined by the reaction terms.

In real applications, however, networks are not symmetric or, as they are called, undirected. It is often the case that a connection between two distinct nodes is associated with a specific direction, which means that the resulting graph is directed. Think for instance of the human mobility flow patterns with their immediate consequences for urban planning, transportation design and epidemic control. Several routes can be crossed in one direction only, so losing the symmetry between pairs of nodes. The internet and the cyberworlds, networks that we explore on a daily basis, are also characterised by an asymmetric routing of the links. Traffic symmetry typically does not hold for network locations beyond intranet and access links \cite{john}. Furthermore, the map of the neural connection in the brain is also asymmetric, due to the physiology of neurons \cite{neuron2}. This can be seen in connectome models \cite{neuron4,connectome,neuron3}, coarse grained maps of the brain, revealed by MRI experiments, which reflects this asymmetric arrangement of connections at different scales of resolution. 

Motivated by this observation, we will show that topology-driven instabilities can develop for reaction-diffusion systems on directed graphs, even when the examined system cannot experience a Turing-like or wave instability if defined on a regular lattice. This is at odds with the behaviour of symmetric networks. Therefore, the characteristics of the spatial support that hosts the interacting species play a role of paramount importance, often neglected and under appreciated in the literature. The obvious consequence of our analysis is a paradigmatic shift in the conventional approach to dynamical modelling of any given natural phenomenon. The dynamical rules of interactions, on which the attention of the modellers has been so far solely devoted, are not the only instigators in the complex interactions that yield the rich variety of self-organised patterns seen in real-world applications. The topology of the space is also important, and significantly influences the conditions that yield the dynamical instability. This novel perspective might also hint at the causes that seem to favour an evolutionary selection of network-like structures in the organisation of living matter at different scales. For instance, the excitatory and inhibitory dynamics of neurons, and their specific spatial arrangement are two ingredients that cannot be trivially decoupled, the first being probably optimised to serve the global functioning of the brain, given the peculiar structure of the second.  

The paper is organised as follows. In Section~\ref{sec:results} we will start by briefly discussing the mathematics of reaction-diffusion on a network, to provide some background for what is to follow. 
Then, we will turn to developing the generalised theory of patterns formation on a directed network. We will test the 
adequacy of the theory, for specific reference models, for different choices of the underlying networks and employing two plausible 
definitions of the Laplacian operator. Travelling waves and stationary inhomogeneous 
patterns will be obtained, by changing the topological characteristic of the graphs, while fixing the parameters to values for which the patterns cannot emerge when the model is defined on a symmetric spatial support. In Section~\ref{sec:concl} we summarise our results and draw conclusions. Section~\ref{sec:methods} contains technical details not included in the earlier presentation.

\section{Results}
\label{sec:results}

\subsection*{Background}

We begin by considering a network made of $\Omega$ nodes and characterised by the $\Omega \times \Omega$ adjacency matrix $\mathbf{W}$: the entry 
$W_{ij}$ is equal to one if nodes $i$ and $j$ (with $i \ne j$) are connected, and it is zero otherwise. If the network is undirected, the matrix
$\mathbf{W}$ is symmetric. Consider then two interacting species and label with $(\phi_i, \psi_i)$ their respective concentrations on node $i$. A general reaction-diffusion system defined on the network takes the form:

\begin{eqnarray}
\frac{d \phi_i}{d t} &=& f(\phi_i,\psi_i)+ D_{\phi}\sum_j \Delta_{ij} \phi_j \nonumber\\
\frac{d \psi_i}{d t} &=& g(\phi_i,\psi_i)+D_{\psi} \sum_j \Delta_{ij} \psi_j,
\label{eq:reac_dif}
\end{eqnarray}
where $D_{\phi}$ and $D_{\psi}$ denote the diffusion coefficients of, respectively, species $\phi$ and $\psi$; $f(\cdot, \cdot)$ and $g(\cdot, \cdot)$ 
are nonlinear functions of the concentrations and stand for the reaction terms; $\Delta_{ij}=W_{ij}-k_i \delta_{ij}$ is the network Laplacian, where $k_i$ refers to the connectivity of node $i$ and $\delta_{ij}$ is the Kronecker's delta.  Imagine now that system (\ref{eq:reac_dif}) admits an homogeneous fixed point, which we label $(\phi^*,\psi^*)$. This implies $f(\phi^*,\psi^*)=g(\phi^*,\psi^*)=0$. We also require that the 
fixed point is stable, namely that $\textrm{tr}(\textbf{J})=f_{\phi}+g_{\psi}<0$ and $\textrm{det}(\textbf{J})=f_{\phi} g_{\psi}-f_{\psi} g_{\phi}>0$, where 
$\textbf{J}$ is the Jacobian of (\ref{eq:reac_dif}) evaluated at $(\phi^*,\psi^*)$ and $f_{\phi}$ (resp. $f_{\psi}$) denotes the derivatives of $f$ with respect to $\phi$ (resp. $\psi$), and similarly for $g_{\phi}$ and $g_{\psi}$.
 
Patterns arise when $(\phi^*,\psi^*)$ becomes unstable with respect to inhomogeneous perturbations. To look for instabilities, one can introduce a small perturbation ($\delta \phi_i$, $\delta \phi_i$) to the fixed point and linearise around it. In doing so, one obtains Eq.~(\ref{eq:linear}) (see Methods).
For regular lattices, the Fourier transform is usually employed to solve this linear system of equations. 
When the system is instead defined on a network, a different procedure needs to be followed \cite{nakao,asllani}. To this end, we define the eigenvalues and eigenvectors of the Laplacian operator:
\begin{equation} \label{eq:lapegv}
\sum_{j=1}^{\Omega} \Delta_{ij} v^{(\alpha)}_j =  \Lambda^{(\alpha)} v^{(\alpha)}_i, \quad \alpha = 1,\ldots,\Omega. 
\end{equation}  
When the network is undirected, the Laplacian is symmetric. Therefore the eigenvalues $\Lambda^{(\alpha)}$ are real and the eigenvectors $v^{(\alpha)}$ form an orthonormal basis. This condition needs to be relaxed when dealing with the more general setting of a directed graph, which we will discuss in the next subsection. The inhomogeneous perturbations $\delta \phi_i$ and $\delta \psi_i$ can be expanded as:

\begin{equation}
\delta \phi_i=\sum_{\alpha=1}^{\Omega}c_{\alpha}e^{\lambda_\alpha t}v^{(\alpha)}_i,\hspace*{1cm}
\delta \psi_i=\sum_{\alpha=1}^{\Omega}b_{\alpha}e^{\lambda_\alpha t}v^{(\alpha)}_i.
\label{eq:exp}
\end{equation}
The constants $c_\alpha$ and $b_\alpha$ depend on the initial conditions. By inserting the above expressions in  
Eq. (\ref{eq:linear}) (see Methods) one obtains the dispersion relation 
\begin{equation}
\lambda_{\alpha} = \frac{1}{2}\left(\textrm{tr}\textbf{J}_{\alpha} + \sqrt{(\textrm{tr}\textbf{J}_{\alpha})^2-4\textrm{det}\textbf{J}_{\alpha}}\right),
\label{disp_rel}
\end{equation}
which characterises the response of the system in Eq.~(\ref{eq:reac_dif}) to external perturbations. Here $\textbf{J}_{\alpha}$ is the modified Jacobian matrix, accounting for diffusion contributions. This quantity is defined in the Methods section.

We use $\left(\lambda_{\alpha} \right)_{\textrm{Re}}$ and $\left(\lambda_{\alpha} \right)_{\textrm{Im}}$ to label respectively the real and imaginary parts of $\lambda (\Lambda^{(\alpha)})$. If $\left(\lambda_{\alpha} \right)_{\textrm{Re}} >0$, the fixed point is unstable and the system exhibits a pattern whose spatial properties are encoded in $\Lambda^{(\alpha)}$. These are the celebrated Turing patterns, stable nonhomogeneous spatial motifs, that result from a resonant amplification of an initial perturbation of the homogeneous fixed point. The quantity $\Lambda^{(\alpha)}$ is the analogue of the wavelength for a spatial pattern in a system defined on a continuous regular lattice. In this case $\Lambda^{(\alpha)} \equiv -k^2$, where $k$ labels the usual spatial Fourier frequency.

It can be proved that, for symmetric graphs, $\Lambda^{(\alpha)}$ is negative. 
Hence, $\textrm{tr}\textbf{J}_{\alpha}=\textrm{tr}\textbf{J}+(D_{\phi}+D_{\psi})\Lambda^{(\alpha)}<0$. For the instability to manifest, it is therefore sufficient that 
$\textrm{det}\textbf{J}_{\alpha}=\textrm{det}\textbf{J}+(J_{11}D_{\psi}+J_{22}D_{\phi})\Lambda^{(\alpha)}+D_{\psi}D_{\psi}(\Lambda^{(\alpha)})^2<0$, 
a condition that can be met only if: 

\begin{equation}
\label{condTur}
J_{11} D_{\psi}+J_{22} D_{\phi}>0.
\end{equation}

When the above quantity is negative, the Turing instability on a regular lattice, or on a complex symmetric network, cannot take place.
One can further elaborate on the implications of equation (\ref{condTur}), by recalling that $\textrm{tr} \textbf{J} = J_{11}+J_{22}<0$. Hence, 
the instability can only set in  if $J_{11}$ and $J_{22}$ have opposite signs. Without losing generality, we shall hereafter assume 
$J_{11} >0$, and consequently $J_{22} <0$, with  $|J_{22}| > J_{11}$ so to have $\textrm{tr} \textbf{J}<0$. In practical terms, the first species, whose concentration reads $\phi$, acts as an activator, while the second, $\psi$, is the inhibitor. 

Working in this setting, it is instructive to consider the limiting condition $D_{\psi}=0$: only the activators ($D_{\phi} \ne 0$)  can diffuse. The necessary condition (\ref{condTur}) reduces to  $J_{22} D_{\phi}>0$, which is never satisfied since  $J_{22} <0$. In principle, the instability can instead develop if the inhibitors are mobile, $D_{\psi} \ne 0$, and the activator fixed  $D_{\phi} = 0$, but only if the system is defined over a discrete spatial support \footnote{A simple calculation shows that the dispersion relation (\ref{disp_rel}) is  positive for $|\Lambda^{(\alpha)}|>\textrm{det}\textbf{J}/D_{\psi}J_{11}$. If the spatial support is discrete, only a finite number of eigenmodes can be destabilised. The eigenmode associated to the largest eigenvalue $|\Lambda^{(\alpha)}|$ guides the instability and the patterns can develop. At variance, in the continuum case the instability involves smaller and smaller spatial scales. It is therefore not possible to delimit a finite window in  $k$ for which $\left(\lambda\right)_{\textrm{Re}}(k)$ is found to be positive, and, consequently, the Turing instability is prevented to occur. In conclusion, a two species system where only one species can migrate, cannot undergo a classical Turing instability, if the spatial support is assumed continuum.}. 
\newpage
We end this section with an important remark. Since $\textrm{tr}\textbf{J}_{\alpha}<0$, it is clear from equation (\ref{disp_rel}), that it is not possible to satisfy the condition for the instability $\left(\lambda_{\alpha} \right)_{\textrm{Re}} > 0$,  and have, at the same time, an imaginary component of the dispersion relation, $\left(\lambda_{\alpha} \right)_{\textrm{Im}}$, different from zero. This is instead possible when operating in a generalised setting which accommodates for specific long-range couplings in the reaction terms \cite{biancalani} or when considering at least three mutually interacting species \cite{zhab, asllani}. A system unstable for $\Lambda^{(\alpha)} \ne 0$ and with the corresponding $\left(\lambda _{\alpha}\right)_{\textrm{Im}} \ne 0$ is said to undergo a wave-instability and the emerging patterns have the form of travelling waves \cite{biancalani}. As we will discuss in the following, the general scenario that we have outlined here, changes drastically when the reaction-diffusion system operates on a directed graph, rather that on a symmetric network.

\subsection*{Theory of pattern formation on a directed network}

We now turn to considering the case of a directed graph. In this case the adjacency matrix $\mathbf{W}$ is no longer symmetric and its entries $W_{ij}$, if equal to one, indicate the presence of an edge directed from node $i$ to $j$. Now, $k_i=\sum_{j=1}^\Omega W_{ij}$ refers to the outdegree of node $i$, defined as the number of exiting edges from node $i$. The associated Laplacian operator can be defined as $\Delta_{ij}=W_{ij}-k_i \delta_{ij}$, as it is customarily done in cooperative control applications \cite{SaberMurray}, as e.g. intelligent transportation systems, routing of communications and power grid networks. The spreading of physical or chemical substances, rather than information content, requires imposing mass conservation, which results in a different formulation of the Laplacian operator \cite{jost}, where $W_{ij}$  is formally replaced by $W_{ji}$ in the definition of $\Delta_{ij}$, as it can be readily obtained from a simple microscopic derivation. Clearly, the two operators coincide, when defined on a symmetric network. 
\newpage
For a directed graphs, the eigenvalues of both Laplacian operators, will in general be complex, $\Lambda^{(\alpha)}\in\mathbf{C}$. This property requires the development of a generalised theory of the instabilities, extending the analysis outlined in the previous section. As usual, we will assume a stable homogeneous fixed point \footnote{To carry out the linear stability analysis, the homogeneous fixed point should be solution of the spatially extended system of governing equations. This prescription implies dealing with a balanced network (the outgoing connectivity equals the incoming one), when Fickean diffusion of material entities is being addressed $\Delta_{ij}=W_{ji}-k_i \delta_{ij}$ , while no additional hypothesis on the structure of the spatial support are to be made when the operator $\Delta_{ij}=W_{ij}-k_i \delta_{ij}$ is assumed to hold.}, and indicate with $\textbf{J}$ the associated Jacobian matrix. The stability of the fixed point implies $\textrm{tr}(\textbf{J})<0$ and $\textrm{det}(\textbf{J})>0$. In order to proceed with the linear stability analysis, we must ensure that the eigenvectors are linearly independent. This is not always the case when the underlying graph is directed \footnote{In fact, even without a complete basis, one can use the generalised eigenvectors in order to make progress. In this case however, patterns may emerge, even when the real part of the dispersion relation is negative everywhere, due to a transient growth process \cite{ridolfi}.}. For this reason, the diagonalisability of the Laplacian matrix will be a minimal requirement to satisfy, in our analytical treatment of the stability problem. 

Introducing an inhomogeneous perturbation ($\delta \phi_i, \delta \psi_i$) and linearising around it, one eventually obtains a dispersion relation, formally identical to equation (\ref{disp_rel}), where now $\Lambda^{(\alpha)}$ is a complex quantity. 
For simplicity, and with obvious meaning of the notation,  we write $\Lambda^{(\alpha)}=\Lambda^{(\alpha)}_{\textrm{Re}}  +i \Lambda^{(\alpha)}_{\textrm{Im}}$, where $\Lambda^{(\alpha)}_{\textrm{Re}}<0$, since the Laplacian matrix spectrum falls, according to Gerschgorin circle theorem \cite{ger}, in the left half of the complex plane. A simple algebraic manipulation yields the following preliminary relations:

\begin{eqnarray}
(\textrm{tr}\textbf{J}_{\alpha})_{\textrm{Re}} &=&\textrm{tr}\textbf{J}+(D_{\phi}+D_{\psi}) \Lambda^{(\alpha)}_{\textrm{Re}} <0\nonumber\\
(\textrm{tr}\textbf{J}_{\alpha})_{\textrm{Im}} &=&(D_{\phi}+D_{\psi}) \Lambda^{(\alpha)}_{\textrm{Im}}\nonumber\\
(\textrm{det}\textbf{J}_{\alpha})_{\textrm{Re}}&=&\textrm{det}\textbf{J}+(J_{11}D_{\psi}+J_{22}D_{\phi}) \Lambda^{(\alpha)}_{\textrm{Re}}\nonumber \\&+&D_{\phi}D_{\psi}\left[\left(\Lambda^{(\alpha)}_{\textrm{Re}}\right)^2-\left(\Lambda^{(\alpha)}_{\textrm{Im}}\right)^2 \right]
\label{notation}\\
(\textrm{det}\textbf{J}_{\alpha})_{\textrm{Im}} &=&(J_{11}D_{\psi}+J_{22}D_{\phi})\Lambda^{(\alpha)}_{\textrm{Im}}+2D_{\phi}D_{\psi} \Lambda^{(\alpha)}_{\textrm{Re}} \Lambda^{(\alpha)}_{\textrm{Im}}, \nonumber
\end{eqnarray}
\newpage
\noindent where $(\cdot)_{\textrm{Re}}$ and $(\cdot)_{\textrm{Im}}$ select respectively the real and imaginary parts of the quantity in between brackets. To study the dispersion relation (\ref{disp_rel}), we shall make use of an elementary property  of the square root of a complex number $z=a+bi$, namely that
\begin{equation}
\sqrt{z}=\pm\left(\sqrt{\dfrac{a+|z|}{2}}+ \textrm{sgn}(b)\sqrt{\dfrac{-a+|z|}{2}}i \right),
\end{equation}
where $\textrm{sgn}(\cdot)$ is the standard sign function. In light of this consideration, the dispersion relation can be cast in the form: 

\begin{equation}
\lambda_{\alpha}=\frac{1}{2}\left[ (\textrm{tr}\textbf{J}_{\alpha})_{\textrm{Re}} + \gamma\right]+\frac{1}{2}\left[ (\textrm{tr}\textbf{J}_{\alpha})_{\textrm{Im}} + \delta\right]i,
\label{disp_rel_comp}
\end{equation}
where:
\begin{eqnarray}
\gamma&=&\sqrt{\dfrac{A+\sqrt{A^2+B^2}}{2}}\nonumber\\
\delta&=& \textrm{sgn}(B)\sqrt{\dfrac{-A+\sqrt{A^2+B^2}}{2}},
\end{eqnarray}
and:

\begin{eqnarray}
A&=&\left[(\textrm{tr}\textbf{J}_{\alpha})_{\textrm{Re}}\right]^2-\left[(\textrm{tr}\textbf{J}_{\alpha})_{\textrm{Im}} \right]^2-4(\textrm{det}\textbf{J}_{\alpha})_{\textrm{Re}}, \nonumber\\ 
B&=& 2 (\textrm{tr}\textbf{J}_{\alpha})_{\textrm{Re}} (\textrm{tr}\textbf{J}_{\alpha})_{\textrm{Im}} -4 (\textrm{det}\textbf{J}_{\alpha})_{\textrm{Im}}.
\end{eqnarray}

The dispersion relation can now contain an imaginary contribution, which bears the signature of the imposed network topology. Travelling waves 
can in principle materialise for a two species reaction-diffusion model on a directed network, at variance with what happens when the system is defined on a symmetric spatial support. For the instability to occur,  $\left( \lambda_{\alpha} \right)_{\textrm{Re}}$ must be greater than zero, which is equivalent to setting  $|R_t| \leq \gamma$, as it follows immediately from equation (\ref{disp_rel_comp}). A straightforward, although lengthy, calculation which involves relations (\ref{notation}), allows us to rewrite the condition for the instability in the compact form:

\begin{equation}
S_2(\Lambda^{(\alpha)}_{\textrm{Re}}) \left[\Lambda^{(\alpha)}_{\textrm{Im}}\right]^2 \leq -S_1(\Lambda^{(\alpha)}_{\textrm{Re}}),
\label{instability}
\end{equation}

where $S_1$ and $S_2$ are polynomials in $\Lambda^{(\alpha)}_{\textrm{Re}}$, of fourth and second degree respectively. The coefficients of the polynomials are model dependent, as specified in Eqs.~(\ref{C0}), (\ref{C1}) and (\ref{C2}) in the Methods. Given the model, one can construct a class of graphs, namely those whose spectral properties match condition 
(\ref{instability}), for which the instability takes place. Here we are particularly interested in models that cannot develop the instability when defined on a symmetric, hence undirected, graph. In this case $\Lambda^{(\alpha)}_{\textrm{Im}}=0$, by definition, and the generalised condition of instability (\ref{instability}) cannot be met. On the contrary, when the graph is made directed, $\Lambda^{(\alpha)}_{\textrm{Im}} \ne 0$ and the examined models can experience a topology-driven instability, as governed by relation (\ref{instability}). 

Recall that we are linearising around a stable fixed point, hence $\textrm{tr}\textbf{J}=J_{11} + J_{22}<0$ and, in addition, $\textrm{det}\textbf{J}>0$. Because of the first inequality, the setting with both $J_{11}$ and $J_{22}$ positive is, as expected, ruled out a priori. Consider instead the dual scenario, where both $J_{11}$, $J_{22}$ are taken to be negative. Hence, $S_1$ and $S_2$ are positive, as it can be immediately appreciated by direct inspection of Eqs.~(\ref{C1}) and (\ref{C2}) in the Methods \footnote{The coefficients $C_{11}$, $C_{13}$, $C_{15}$ are negative, while  $C_{12}$, $C_{14}$ are positive. Recalling that $\Lambda^{(\alpha)}_{\textrm{Re}}<0$, it is straightforward to conclude that $S_1$ and $S_2$ are indeed positive quantities.}, and the instability condition (\ref{instability}) cannot be achieved. In conclusion, and in agreement with the standard theory of pattern formation on a regular symmetric lattice (network), $J_{11}$ and $J_{22}$ must have opposite signs, for the instability to develop. 
 
 In analogy with the preceding discussion, we assume that the reaction-diffusion scheme is such that $J_{11} >0$, and $J_{22} <0$. We imagine here that the necessary condition (\ref{condTur}) for the instability to set in, is not satisfied. No patterns can therefore develop, according to the classical paradigm. 
 Under this assumptions, while  $S_1$ is still a positive definite quantity,  $S_2$ can take negative values, since $C_{20}<0$. The generalised condition (\ref{instability}) can then be satisfied and one can expect topology-driven instabilities to develop on directed graphs, as we shall demonstrate in the forthcoming section. 
 
 As an additional point, we wish to emphasise that patterns on a directed network can also emerge when the inhibitors are prevented from diffusing ($D_{\psi}=0$, in our notation), at odds with the conventional vision. Furthermore, travelling waves are always found for a two species model, 
 the imaginary part of the dispersion relation $\lambda_{\alpha}$ being at all times different from zero, inside the instability domain. However, as we will demonstrate in the next section, when $\left(\lambda_{\alpha}\right)_{\textrm{Im}}<<\left(\lambda_{\alpha}\right)_{\textrm{Re}}$, the system evolves towards a stationary inhomogeneous pattern, reminiscent of the Turing instability. In conclusion, a two species reaction-diffusion system on a directed graph can display a rich variety of instabilities, beyond the conventional scenario which applies to symmetric, hence undirected, spatial supports.  
 
\subsection*{Numerical results}

To challenge the theoretical analysis carried out in the previous section, we here focus on a specific system, the so called Brusselator model, 
a paradigmatic example of autocatalytic reactions. Details of this model are presented in the Methods section. 
It is worth stressing that the Brusselator model has been selected for demonstrative purposes and due to its pedagogical value: the techniques here developed, and the conclusions that we shall reach, are nevertheless general and apply to a wide range of models. As an example, in the appendix we will repeat the same analysis for a version of the Fitzhugh-Nagumo scheme, that serves as a toy model for inspecting the coupled dynamics of neurons. 

We will start by considering the first formulation of the Laplacian operator, $\Delta_{ij}=W_{ij}-k_i \delta_{ij}$, which, as previously mentioned, is employed in cooperative control applications \cite{SaberMurray}. To proceed, we should also specify the characteristics of the networks that define the spatial backbone for the explored systems. Two different strategies for generating the graphs will be considered. On the one side, we will follow the Watts-Strogatz (WS) approach \cite{watts}, modified in order to make the network directed. We will also implement an alternative generation strategy, the so-called Newman-Watts (NW) algorithm \cite{newman1}. Both approaches are described in the Methods section. When $\Delta_{ij}=W_{ji}-k_i \delta_{ij}$ is assumed instead, the directed networks are created with the additional prescription to balance the number of incoming and outgoing connections per node. 

In the following we will report the emergence of topology-driven instabilities, defined on directed networks assembled via the two strategies mentioned here. 
\begin{figure*}
\begin{tabular}{cccc}
\includegraphics[scale=0.85]{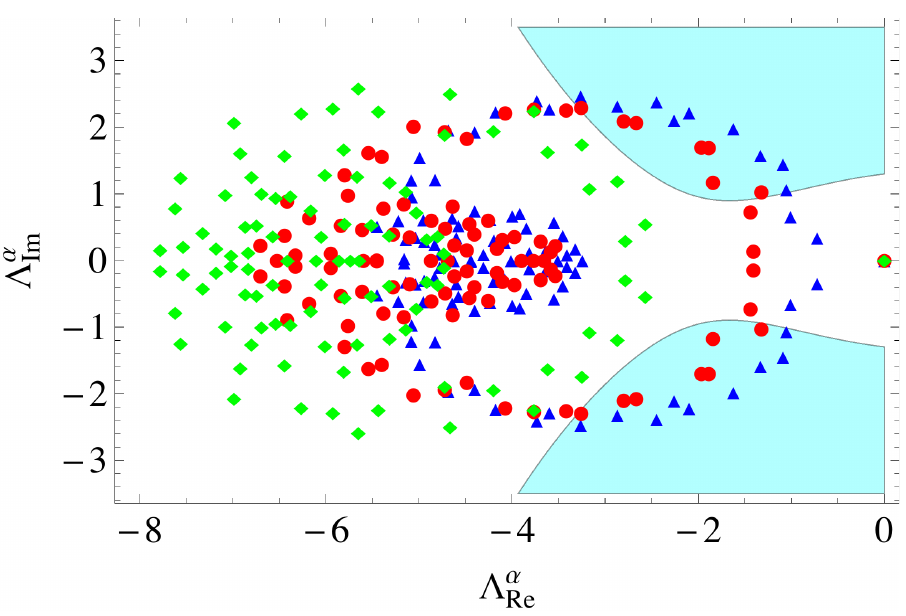}&\hspace{0.5em}
\includegraphics[scale=0.47]{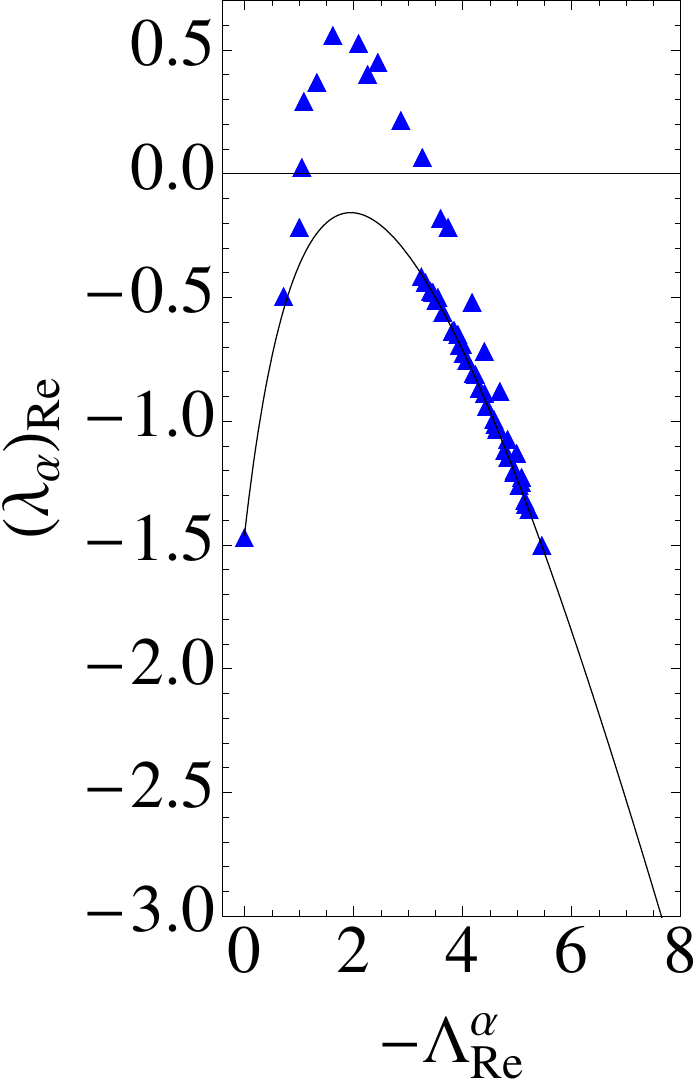}&\hspace{-2.5em}
\includegraphics[scale=0.47]{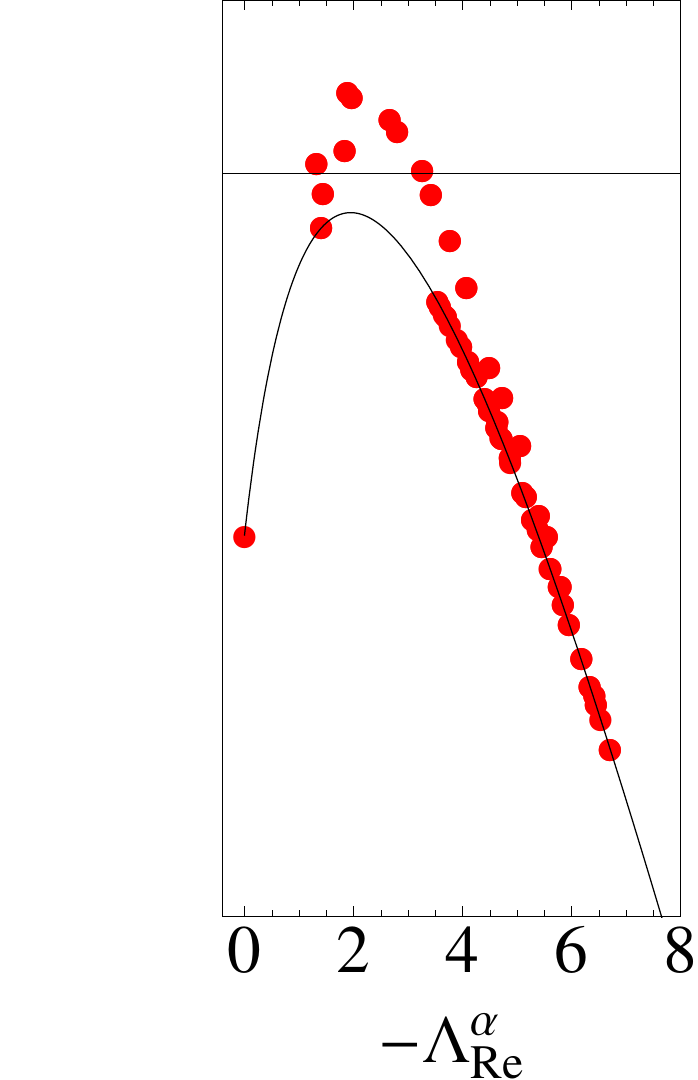}&\hspace{-2.5em}
\includegraphics[scale=0.47]{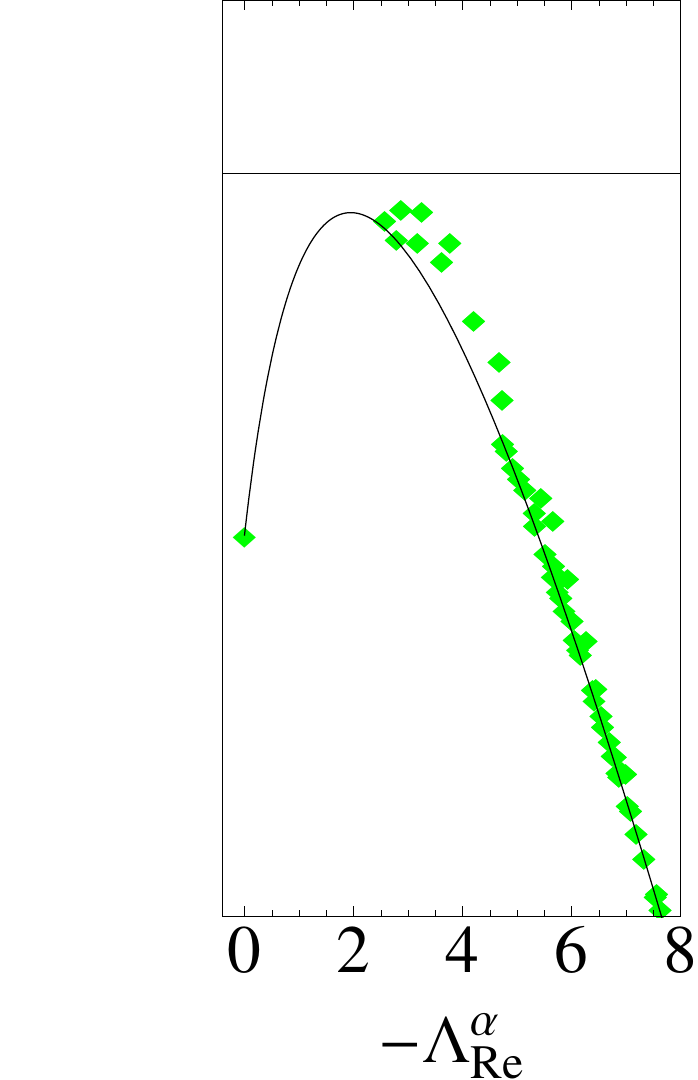}\\
(a) & & (b) &\\
\end{tabular}
\caption{Panel (a): Spectral plot of three Laplacians generated from the Newman-Watts algorithm for $p=0.27$, $p=0.5$ and $p=0.95$ (blue triangles, red circles and green diamonds respectively) and network size $N=100$. The shaded area indicates the instability region for the case of the Brusselator model, where the parameters are $b=9$, $c=30$, $D_{\phi}=1$ and $D_{\psi}=7$.
Panel (b): The real part of the dispersion relation for the same three choices of Newman-Watts networks as in panel (a).  The black line originates from the continuous theory. 
}
\label{figure1}
\end{figure*}

\vspace{0.5cm}
\textbf{1. Travelling waves for systems of two diffusing species.} We commence by considering the Brusselator model in the general setting with $D_{\phi} \ne 0$ and $D_{\psi} \ne 0$. That is, both the 
activators and inhibitors are allowed to diffuse between connected nodes of the network. Furthermore, we set the parameters so that 
the system is stable versus external perturbations of the homogeneous fixed point 
and no patterns can develop if the spatial support is assumed symmetric (or continuous). The generalised condition (\ref{instability}) for the instability on a  directed network can be graphically illustrated in the reference plan $(\Lambda^{(\alpha)}_{\textrm{Re}}, \Lambda^{(\alpha)}_{\textrm{Im}})$. Given the parameters of the model, one can in fact estimate the coefficients $C_{1q}$ ($q=0,..,4$) and $C_{2q}$ ($q=0,1,2$) via Eqs.~(\ref{C1}) and (\ref{C2}) in the Methods. Then, the inequality (\ref{instability}) allows one to delimit a model-dependent region of instability, which is depicted with a shaded area in Figure (\ref{figure1}), panel (a). Each eigenvalue of the discrete Laplacian (defined on the directed network) appears as a point in the plane $(\Lambda^{(\alpha)}_{\textrm{Re}}, \Lambda^{(\alpha)}_{\textrm{Im}})$. If a subset of the $\Omega$ points that define the spectrum of the Laplacian fall inside the region outlined above, the instability can take place. For an undirected graph, $\Lambda^{(\alpha)}_{\textrm{Im}}=0$, and the points are located on the real (horizontal) axis, thus outside the instability domain.   
In panel (a) of Figure~\ref{figure1} the spectral plot of three Laplacians, generated from the NW algorithm, are displayed for three choices of $p$. As $p$ increases the points move to the left, away from the region of instability. The regular ring that captures the eigenvalue distribution for low $p$ (blue triangles), becomes progressively distorted: for $p=0.95$ the points (green diamonds) partially fill a circular patch in  ($\Lambda^{(\alpha)}_{\textrm{Re}}, \Lambda^{(\alpha)}_{\textrm{Im}}$). In panel (b) of Figure~\ref{figure1} the real part $\left(\lambda_{\alpha}\right)_{\textrm{Re}}$ of the dispersion relation is plotted as a function of $-\Lambda^{(\alpha)}_{\textrm{Re}}$. The black solid line refers to the continuous theory: no instability can develop in the limit of continuum space, since $\left(\lambda_{\alpha}\right)_{\textrm{Re}}$ is always negative. However, when the Brusselator model evolves on a discrete support, the continuous line is replaced by a collection of $\Omega$ points. When the instability condition (\ref{instability}) is fulfilled, the points are lifted above the solid curve and cross the horizontal axis. The figure shows that $\left(\lambda_{\alpha}\right)_{\textrm{Re}}$ is positive over a finite domain in $-\Lambda^{(\alpha)}_{\textrm{Re}}$ and the system becomes unstable due to the topology of the underlying directed network. The travelling waves that stem from this instability are displayed in Figure~\ref{waves1}, and also in Supplementary~Movie~1. As we shall prove in the following, inhomogeneous stationary patterns are another possible outcome of the topology-driven instability.

\begin{figure}
\includegraphics[scale=0.36]{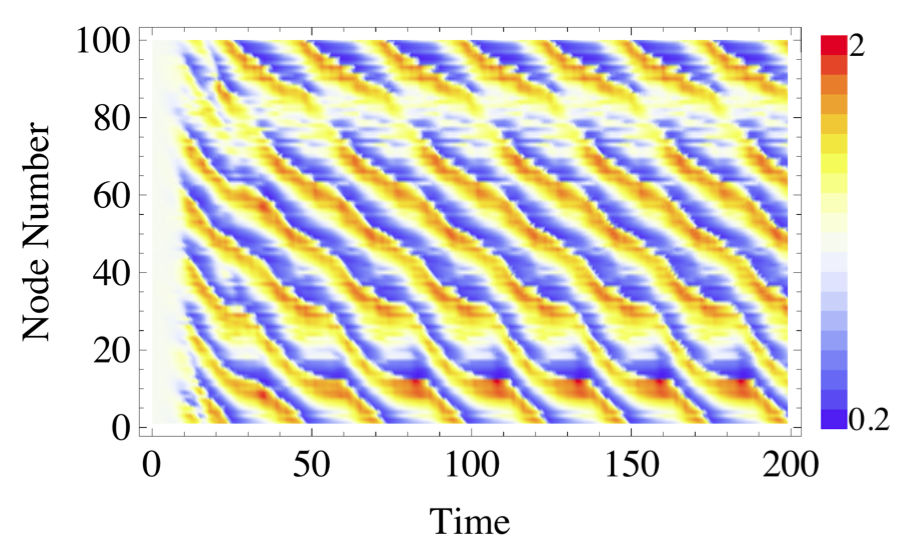}
\caption{Time series for the case of the Brusselator model on a Newman-Watts network, generated with $p=0.27$. The nodes are ordered as per the original lattice. Details of the network's spectra and the system's instability are displayed by the blue, triangular symbols in Figure~\ref{figure1}. The caption of that figure also contains the values of the reaction parameters.}
\label{waves1}
\end{figure}

\begin{figure*}
\begin{tabular}{cccc}
\includegraphics[scale=0.85]{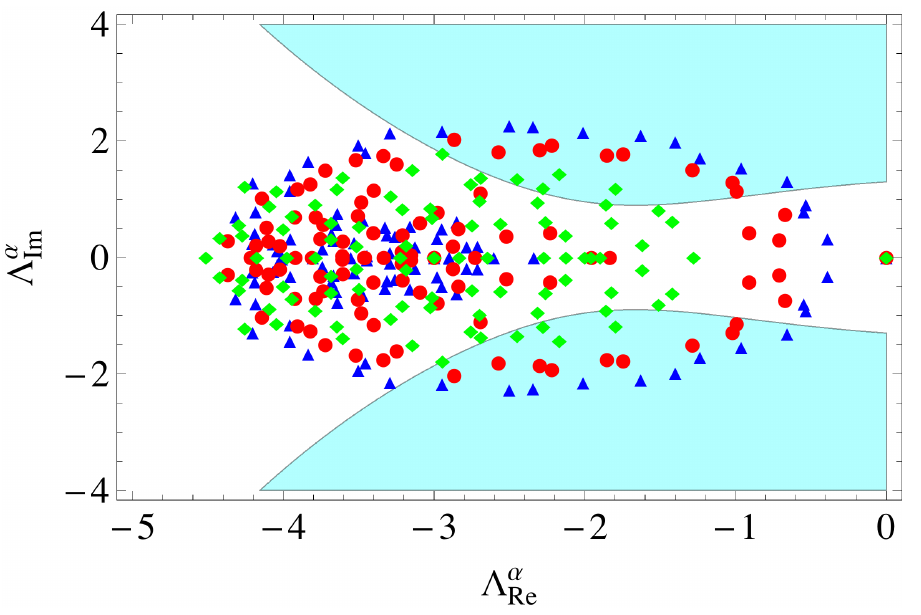}&\hspace{0.5em}
\includegraphics[scale=0.47]{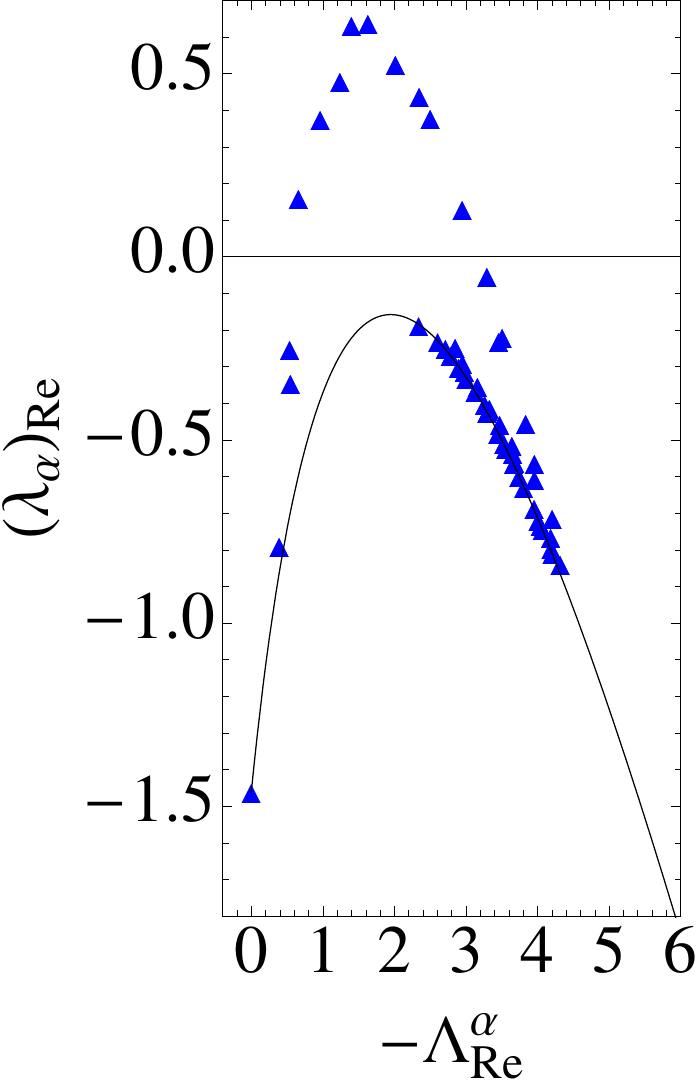}&\hspace{-2.5em}
\includegraphics[scale=0.47]{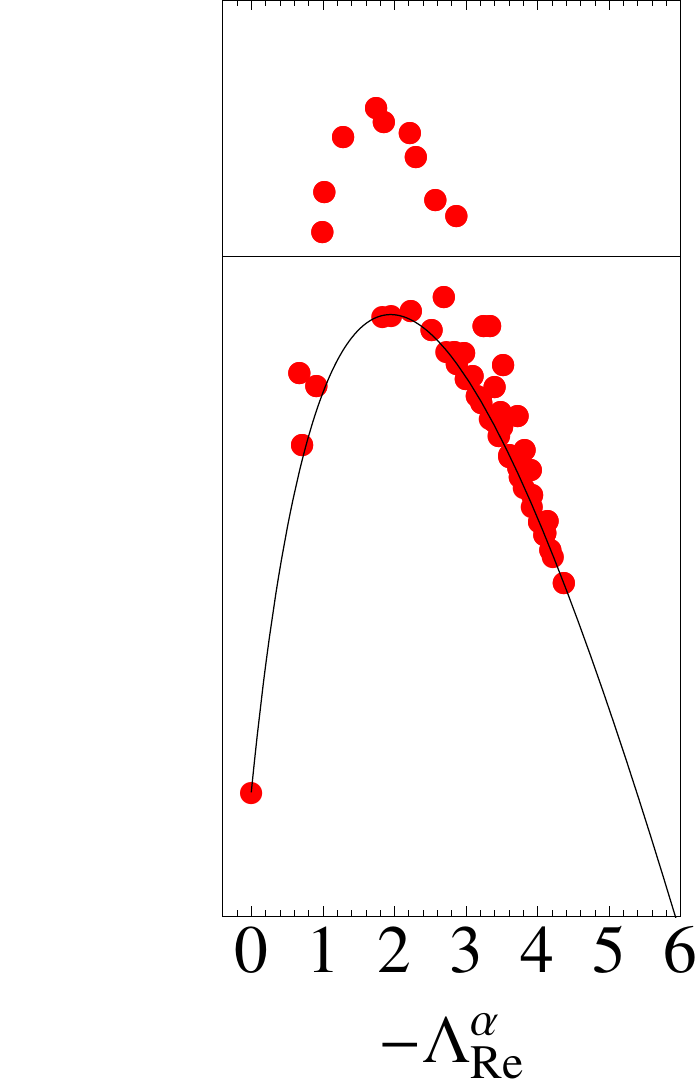}&\hspace{-2.5em}
\includegraphics[scale=0.47]{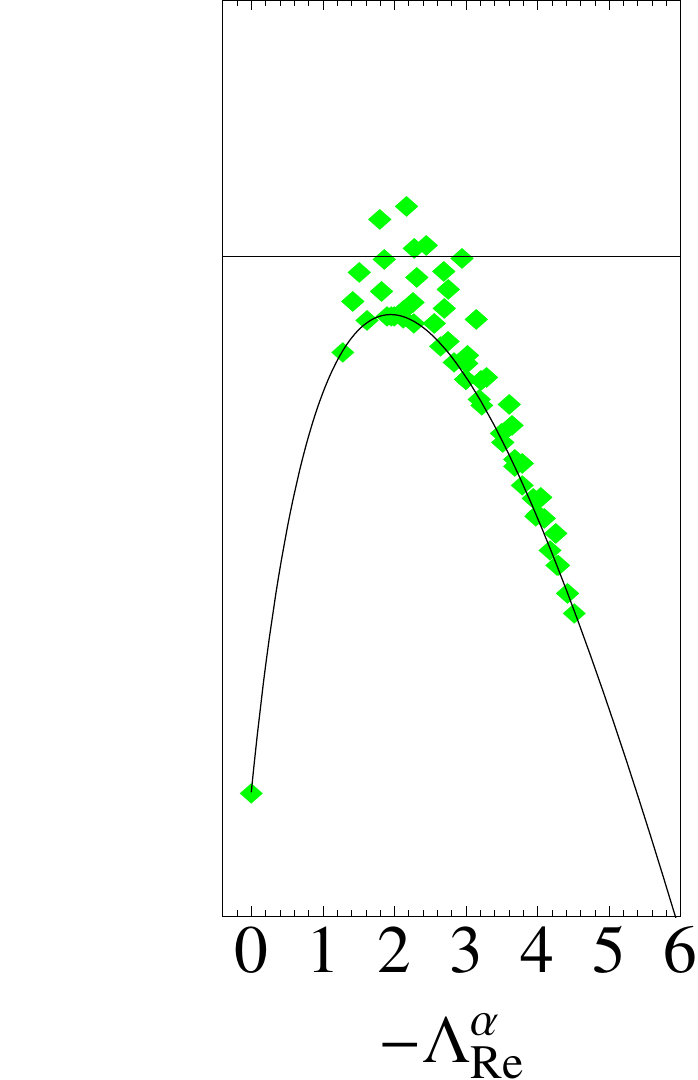}\\
(a) & & (b) &\\
\end{tabular}
\caption{Panel (a): Spectral plot of three Laplacians generated from the Watts-Strogatz method for $p=0.1$, $p=0.2$ and $p=0.8$ (blue triangles, red circles and green diamonds respectively). In all cases the network size is $N=100$. The coloured area indicates the instability region for the Brusselator model. Panel (b): The real part of the dispersion relation for three choices of Watt-Strogatz networks for  $p=0.1$, $p=0.2$ and $p=0.8$ (blue triangles, red circles and green diamonds respectively) and network size $N=100$. 
The parameters are as in Figure \ref{figure1}.}
\label{figure2}
\end{figure*}

\begin{figure}
\includegraphics[scale=0.36]{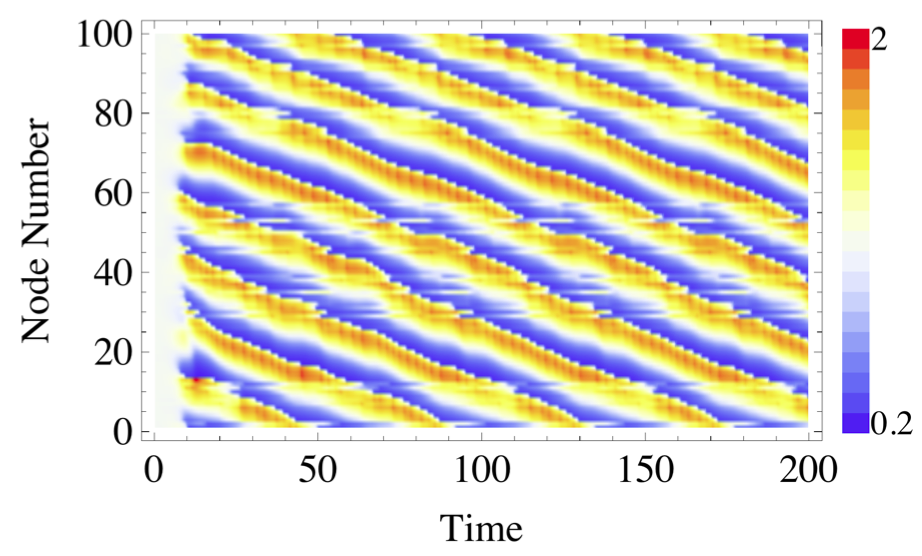}
\caption{Time series for the case of the Brusselator model on a Watts-Strogatz network, generated with $p=0.1$. The nodes are ordered as per the original lattice. Details of the network's spectra and the system's instability are displayed by the blue, triangular symbols in Figure~\ref{figure2}. The reaction parameters are as in Figure \ref{figure1}.}
\label{waves2}
\end{figure}

A similar analysis can be performed when the directed network is generated using the WS method. In panel (a) of Figure \ref{figure2}, the region of instability, outlined by the shaded area, is identical to the one depicted in Figure \ref{figure1}, since the parameters of the reaction-diffusion scheme are unchanged. Here the topology-driven instability occurs for relatively large values of the parameter $p$, when the random nature of the network takes over its small world character. The travelling waves for this system are displayed in Figure~\ref{waves2}.

\begin{figure*}
\begin{tabular}{cc}
\includegraphics[scale=0.85]{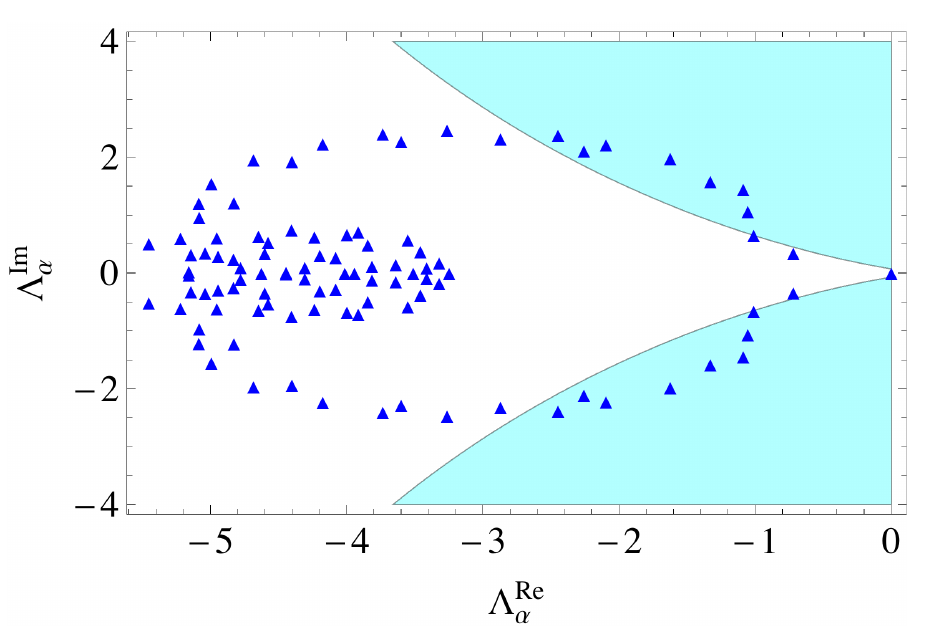}&
\includegraphics[scale=0.85]{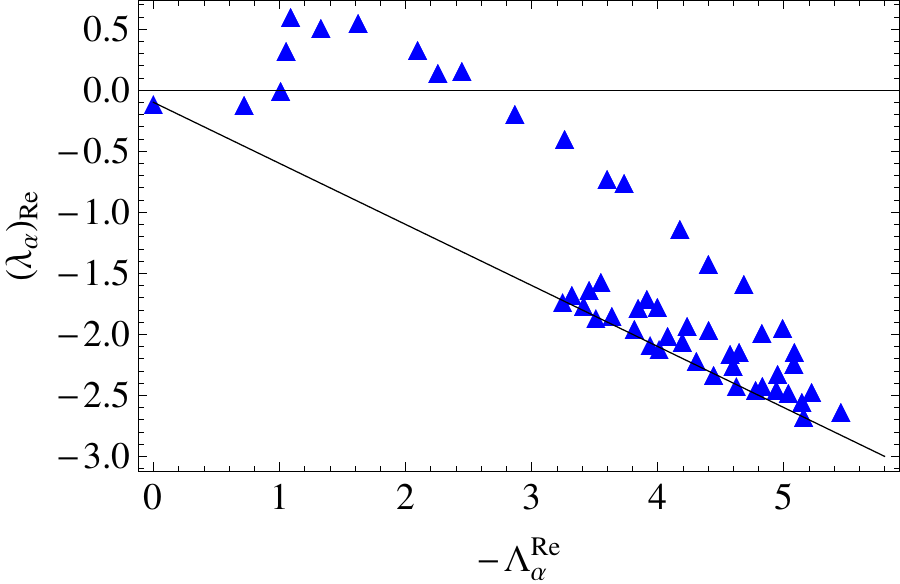}\\
(a) & (b)\\
\end{tabular}
\caption{Panel (a): Spectral plot for a Newman-Watts network with $p=0.27$ and $N=100$. The instability region is plotted for the choice of the Brusselator model, where the diffusion $D_{\psi}$ has been set to zero. Panel (b): The real part of the dispersion relation for a Newman-Watts network with $N=100$, generated with $p=0.27$. The black line is found from the continuous theory. The full parameter set is:  $b=9$, $c=8.2$, $D_{\phi}=1$ and $D_{\psi}=0$.}
\label{figure3}
\end{figure*}

\begin{figure}
\includegraphics[scale=0.36]{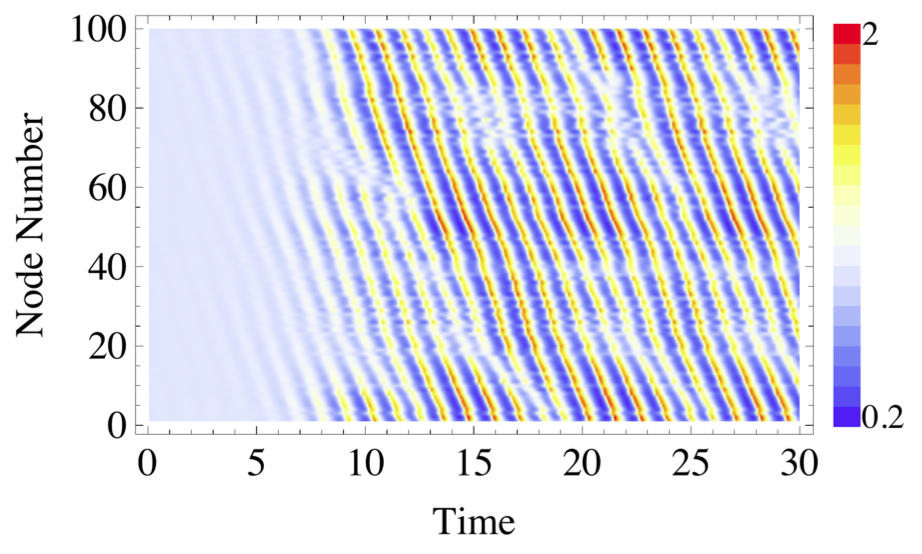}
\caption{Time series obtained from the Brusselator model on a Newman-Watts network, in the case where $D_{\psi}=0$. The network was constructed using $p=0.27$. Details of the network's spectra and the system's instability are displayed by the blue, triangular symbols in Figure~\ref{figure3}. The caption of that figure also contains the values of the reaction parameters.}
\label{waves3}
\end{figure}

\vspace{0.5cm}
\textbf{2. The case of immobile inhibitors, $D_{\psi}=0$.} As mentioned earlier, patterns can also emerge on a directed network when the inhibitors are prevented from diffusing ($D_{\psi}=0$). This is a marginal condition for which classical patterns (both in the continuum limit or on an undirected heterogeneous support) are not found. In panel (a) of Figure \ref{figure3} the condition of instability is represented. The symbols (blue triangles) refer to a NW graph with $p=0.27$ and fall inside the shaded domain, signalling the existence of a topology-driven instability. The same conclusion can be reached upon inspection of panel (b), where the real part of the dispersion relation is plotted and shown to be positive over finite window in $\Lambda^{(\alpha)}_{\textrm{Re}}$. The travelling waves found in this case are shown in Figure~\ref{waves3}.

\begin{figure*}
\begin{tabular}{cc}
\includegraphics[scale=0.8]{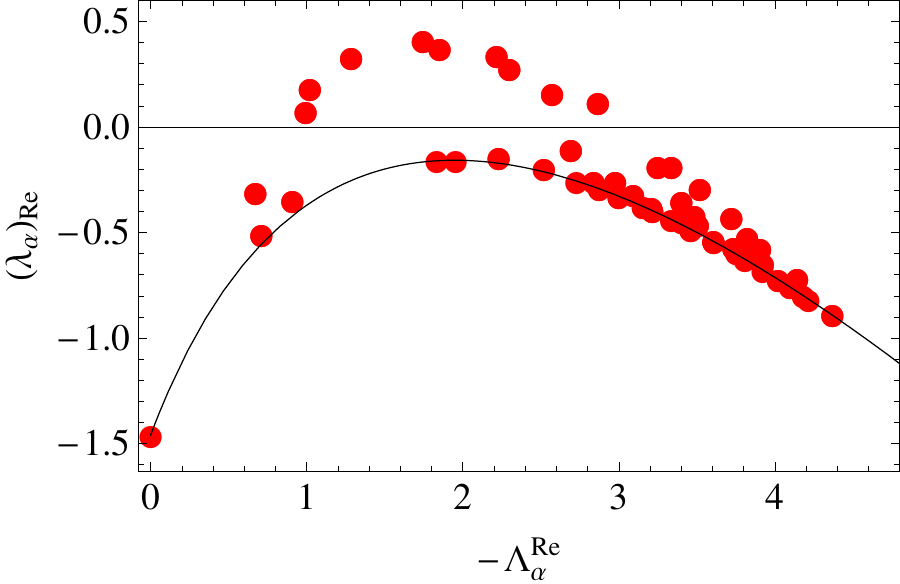}&
\includegraphics[scale=0.8]{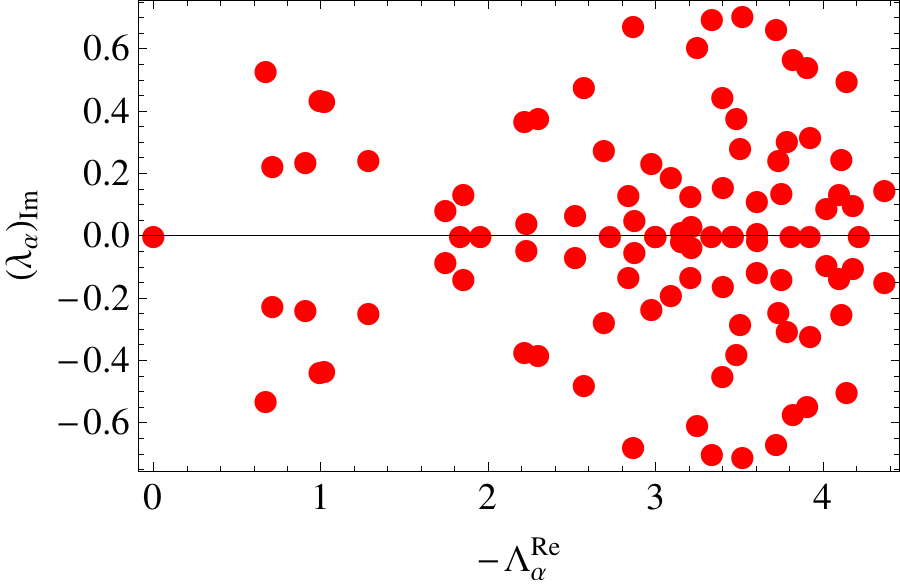}\\
(a) & (b)\\
\end{tabular}
\caption{Panel (a): The real part of the dispersion relation for a Watts-Strogatz network with a Brusselator dynamic. The network is of size $N=100$, generated with $p=0.2$. Panel (b): The imaginary part of the dispersion relation for a Watts-Strogatz network with a Brusselator dynamic. The network is of size $N=100$, generated with $p=0.2$. The parameters are set as in Figure~\ref{figure1}. }
\label{figure4}
\end{figure*}

\begin{figure*}
\begin{tabular}{cc}
\includegraphics[scale=0.24]{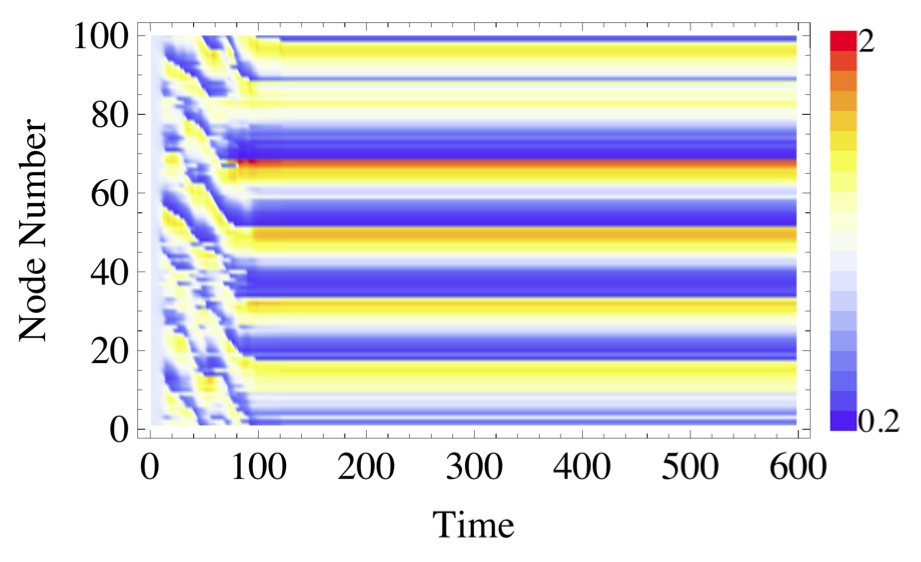}&
\includegraphics[scale=0.77]{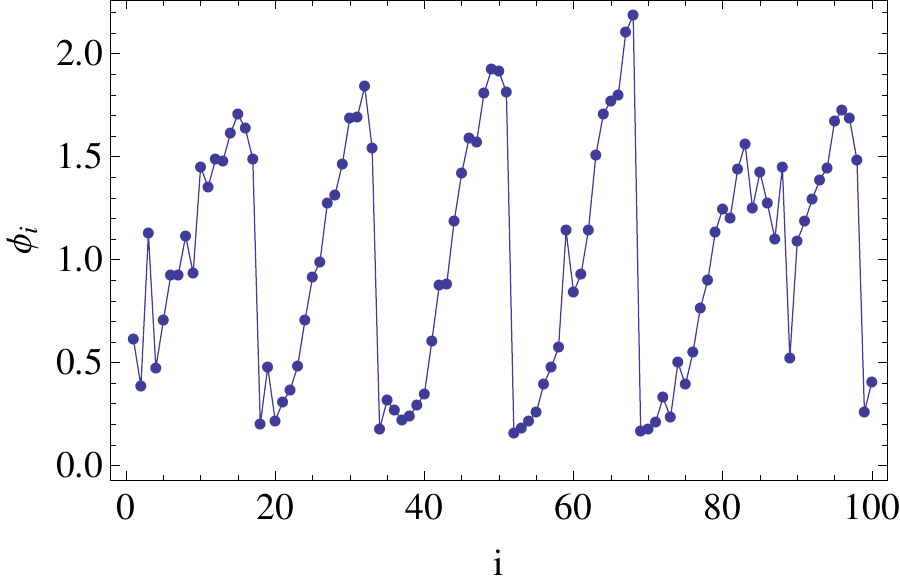}\\
(a) & (b)\\
\end{tabular}
\caption{Panel (a): Time series showing quasi-Turing patterns on a Watts-Strogatz network with a Brusselator dynamic. The network was generated using $p=0.2$. The system evolves from a homogeneous fixed point towards a static pattern. Panel (b); The long-time behaviour of the network, showing the concentration of each node. The dispersion relation for this system is presented in Figure~\ref{figure4}.}
\label{waves_QT}
\end{figure*}

\vspace{0.5cm}
\textbf{3. Stationary inhomogeneous patterns.} The instability mechanism that we have here discussed results in travelling waves, which spread over the direct network. This is due to the fact  
that the imaginary part of the dispersion relation $\lambda_{\alpha}$ is always different from zero, inside the instability domain. Hence, Turing-like 
instabilities, which require imposing $\left(\lambda_{\alpha}\right)_{\textrm{Im}}=0$, cannot formally develop. On the other hand, as we argued earlier,  
stationary inhomogeneous patterns reminiscent of the Turing instability, could be observed, provided $\left(\lambda_{\alpha}\right)_{\textrm{Im}}<<\left(\lambda_{\alpha}\right)_{\textrm{Re}}$. This is the case considered here, as shown in Figure~\ref{figure4}, for a directed network generated via the WS recipe. The parameters of the Brusselator model are instead set as in Figure~\ref{figure1}. This is to stress again that different types of patterns can emerge because of the distinct characteristics of the networks: the patterns are not just selected by the imposed dynamical rules. 
The most unstable mode is located at around $\Lambda^{(\alpha)}_{\textrm{Re}}=-2$ (left panel of Figure~\ref{figure4}), and its corresponding value of   
$\left(\lambda_{\alpha}\right)_{\textrm{Im}}$ is relatively small (right panel of Figure~\ref{figure4}). The patterns found in this situation are shown in the two panels of Figure~\ref{waves_QT}. By comparing Figures~\ref{waves2} (WS with $p=0.1$) and \ref{waves_QT} (WS with $p=0.2$), it is immediately clear that a  
transitions take place, from traveling waves to asymptotically stationary stable, by tuning the rewiring probability $p$ in the WS graph generation scheme.

\vspace{0.1cm}
\textbf{4. Alternative formulation of the transport operator}

\begin{figure*}
\begin{tabular}{cc}
\includegraphics[scale=0.85]{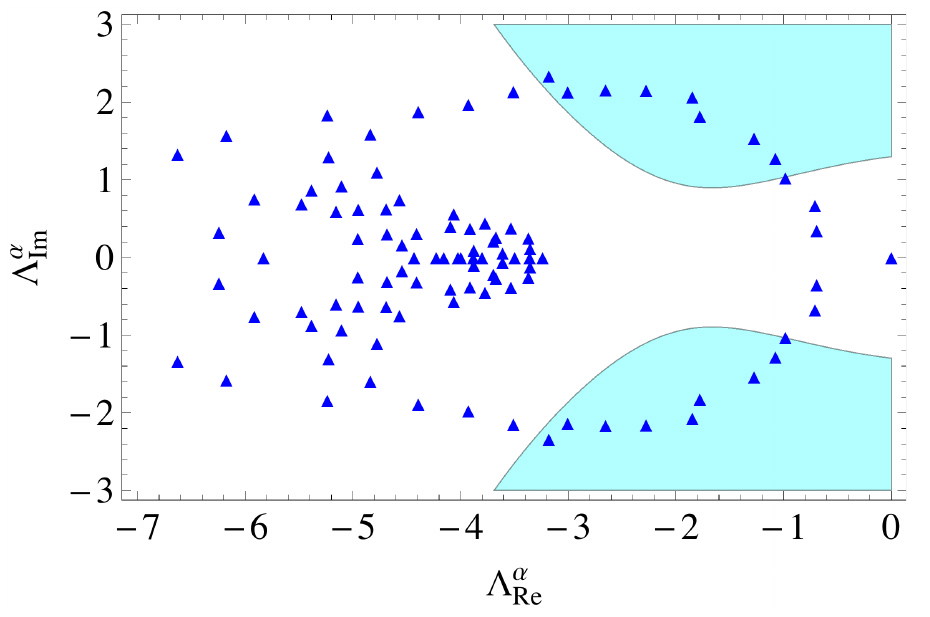}&
\includegraphics[scale=0.85]{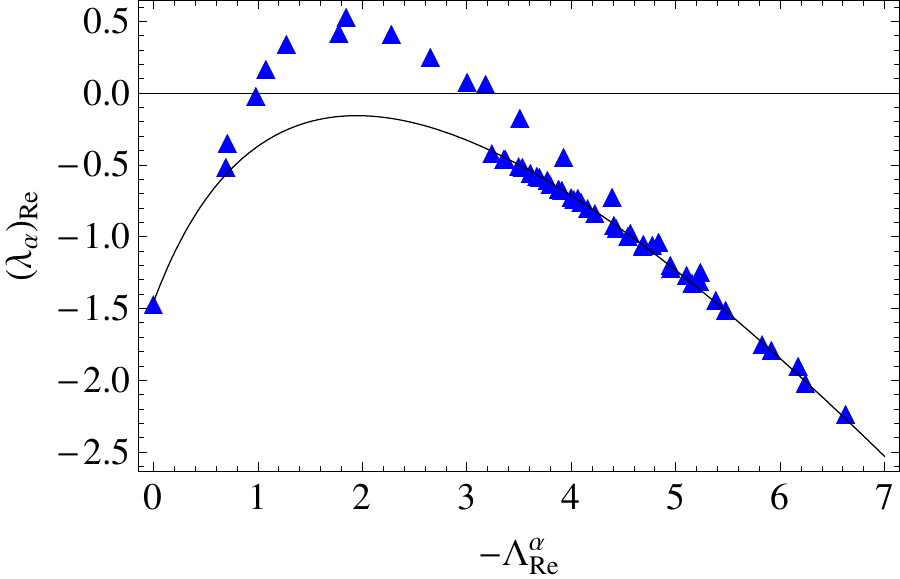}\\
(a) & (b)\\
\end{tabular}
\caption{Panel (a): Spectral plot for a Newman-Watts network with $p=0.27$ and $N=100$. The instability region is plotted for the Brusselator model. Panel (b): The real part of the dispersion relation for a balanced Newman-Watts network with $N=100$, generated with $p=0.27$. Here, $\Delta_{ij}=W_{ji}-k_i \delta_{ij}$. The black line is found from the continuous theory. The full parameter set is:  $b=9$, $c=30$, $D_{\phi}=1$ and $D_{\psi}=7$.}
\label{NW_spectra_new}
\end{figure*}

\begin{figure}
\includegraphics[scale=0.36]{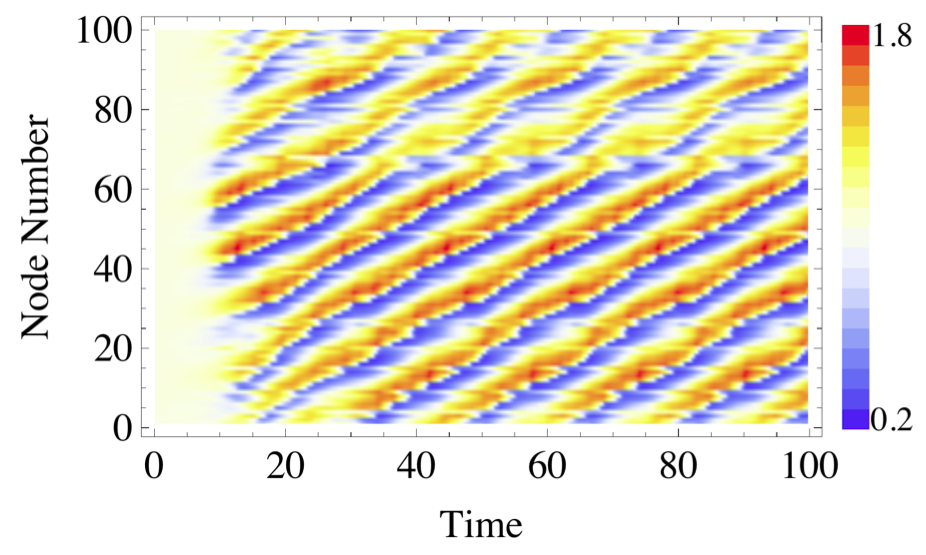}
\caption{Time series obtained from the Brusselator model on a balanced Newman-Watts network, constructed using $p=0.27$. Details of the network's spectra and the system's instability are displayed by the blue, triangular symbols in Figure~\ref{NW_spectra_new}. The caption of that figure also contains the values of the reaction parameters.}
\label{waves_nw_new}
\end{figure}

Consider now the Laplacian operator defined as $\Delta_{ij}=W_{ji}-k_i \delta_{ij}$. As anticipated, this is the operator to be used when interested 
in modelling the diffusive spreading of material entities (molecules, animals) on a network.  To match the assumptions of the analysis, the underlying network has to be balanced: the number of incoming connections ($\sum_j W_{ji}$) needs to be equal to the number of outgoing links ($k_i$). The 
homogeneous fixed point ($\phi^*, \psi^*$) is hence solution of the spatially extended system (\ref{eq:reac_dif}), since 
$\sum_j \Delta_{ij} \phi^*=\sum_j \Delta_{ij} \psi^*=0$, and the linear stability analysis, as discussed above, still holds \footnote{The case of unbalanced networks could be also addressed, provided one linearizes the equations (\ref{eq:reac_dif}) around a non homogeneous state \cite{nonhom}. Such generalization will be discussed elsewhere, as we here aim to a pedagogical introduction to the novel class of topology driven instabilities.}. 
In Figure \ref{NW_spectra_new}, the condition of instability are displayed for a balanced NW network (see Method), while the emerging traveling waves can be clearly appreciated in Figure \ref{waves_nw_new} .

\section{Discussion}
\label{sec:concl}

Patterns can spontaneously develop in reaction-diffusion systems following
a linear instability mechanism, first discussed by Turing in his seminal
paper \cite{turing}. Turing patterns are spontaneously emerging, stationary
inhomogeneous motifs and represent a characteristic form of
ecological self-organisation. Travelling waves can also develop in
reaction-diffusion systems, following a similar instability mechanism. 
Alongside the clear theoretical interest, these concepts connect with many fields of application, from life science to chemistry and physics, transcending the boundaries of ecology.
In general, the inspected model is assumed to be defined on a continuous spatial support or
on a regular lattice.

In many cases, it is instead more natural to establish the system on a
complex network. With reference to ecology,
the nodes of the networks define localised habitat patches, and the
dispersal connection among habitats result in the diffusive coupling
between adjacent nodes. Broadening the discussion, the brain is a
network of neuronal connections, which provide the backbone for the
propagation of the cortical activity. The internet and the cyberword in
general are other, quite obvious examples of applications that require
invoking the concept of network.  In a recent  paper \cite{nakao}, Nakao
and Mikhailov developed the theory of Turing pattern formation on random
symmetric networks, highlighting the peculiarities that stem from the
embedding graph structure. Travelling waves can also set in following an
analogous mechanism \cite{asllani}.

Starting from this context, and to reach a number of potential
applications, we have here considered the extension
of the analysis in  \cite{nakao} to the case of directed, hence non-symmetric, networks.
It is often the case that links joining two distant nodes are defined with an associated direction:
the reactants can move from one node to another, but the reverse action is formally impeded.  In this paper, we have shown that a novel class of
instability, here termed `topology driven', can develop for
reaction-diffusion systems on directed graphs, also when the examined
system cannot experience a Turing-like or wave instability when defined on
a regular lattice
or, equivalently, on a continuous spatial support. This is at variance
with the case of symmetric networks, that cannot possess the intrinsic
ability of turning unstable an otherwise stable homogenous fixed point.
We have shown here that different patterns can be generated depending on the
characteristics of the spatial support on which the reaction-diffusion
system is defined. In particular, transitions from traveling waves to asymptotically
stationary stable patterns, reminiscent of the Turing instability, are obtained when tuning the rewiring probability $p$ in a WS network.
The condition for the instability results in a rather compact mathematical
criterion, which accounts for the spectral properties of the underlying
network and whose predictive adequacy has been validated with reference to
selected case studies.

The existence of a generalised class of instabilities, seeded by the
topological characteristics of the embedding support, suggests a
shift in the conventional approach to modelling of dynamical
systems. Mutual rules of interactions, that define the reactions
among constituents, are certainly important, although not decisive in
determining the asymptotic fate of the system.  The topology of the
space, when assumed to be a directed random network, also matters and plays
an equally crucial role in the onset of the dynamical instability.
This is a general conclusion that we have cast in rigorous terms,
which can potentially inspire novel avenues of research in all those
domains, from neuroscience to social related applications, where networks
prove essential.

\section{Methods}
\label{sec:methods}

\subsection*{Details of the linear stability analysis}

To look for instabilities in Eq.~(\ref{eq:reac_dif}), one can introduce a small perturbation ($\delta \phi_i$, $\delta \phi_i$) to the fixed point and linearise around it. In formulae one finds
\begin{equation} \label{eq:linear}
	\begin{pmatrix} \delta \dot \phi_i \\ \delta \dot \psi_i \end{pmatrix} = \sum_{j=1}^{\Omega} \left( \textbf{J} \delta_{ij} +   \textbf{D} \Delta_{ij} \right) \cdot \begin{pmatrix} \delta \phi_j \\ \delta \psi_j \end{pmatrix},
\end{equation}
where
\begin{equation} 
 \textbf{D}= \left( \begin{smallmatrix} D_{\phi}&0\\ 0&D_{\psi} \end{smallmatrix} \right).
\end{equation}

Expanding the perturbation on the basis of the eigenvectors of the network Laplacian, (see Eq.~\ref{eq:exp})  one obtains the following eigenvalue problem

\begin{equation}
\textrm{det} \left( \begin{array}{ccc}
f_{\phi} + D_{\phi} \Lambda^{(\alpha)}-\lambda_{\alpha} & f_{\psi}  \\
g_{\phi} & g_{\psi} + D_{\psi}\Lambda^{(\alpha)}-\lambda_{\alpha} \end{array}\right)=0.
\label{eq:eig_problem}
\end{equation}
This is equivalent to
\begin{equation}
\textrm{det} \left( \textbf{J}_{\alpha}-\textbf{I}\lambda_{\alpha} \right)=0,
\label{eig_matrix}
\end{equation}
where $\textbf{J}_{\alpha} \equiv \textbf{J}+\textbf{D}\Lambda^{(\alpha)}$ and $\textbf{I}$ stands for the identity matrix.
The eigenvalue with the largest real part, $\lambda_{\alpha}\equiv\lambda_{\alpha}(\Lambda^{(\alpha)})$ defines the dispersion relation presented in Eq.~(\ref{disp_rel}).

\subsection*{Specification of the instability region}

Here we  provide details of the functions $S_1$ and $S_2$, introduced in Eq.~(\ref{instability}) of the paper, using the quantities defined in Eq.~(\ref{notation}). We eventually obtain 
\begin{eqnarray}
\label{C0}
S_1(\Lambda^{(\alpha)}_{\textrm{Re}})&=&C_{14} \left[\Lambda^{(\alpha)}_{\textrm{Re}} \right]^4+C_{13} \left[\Lambda^{(\alpha)}_{\textrm{Re}} \right] ^3+C_{12} \left[\Lambda^{(\alpha)}_{\textrm{Re}} \right] ^2+C_{11}\left[\Lambda^{(\alpha)}_{\textrm{Re}} \right]+C_{10}\\ \nonumber
S_2(\Lambda^{(\alpha)}_{\textrm{Re}})&=& C_{22} \left[\Lambda^{(\alpha)}_{\textrm{Re}} \right]^2+C_{21} \left[\Lambda^{(\alpha)}_{\textrm{Re}} \right]+C_{20},\nonumber
\end{eqnarray}
where
\begin{eqnarray}
\label{C1}
C_{14}&=&  D_{\phi}D_{\psi}(D_{\phi}+D_{\psi})^2\nonumber\\
C_{13}&=& (D_{\phi}+D_{\psi})^2(J_{11}D_{\psi}+J_{22}D_{\phi})+2\textrm{tr}\textbf{J}D_{\phi}D_{\psi}(D_{\phi}+D_{\psi})\nonumber\\
C_{12}&=& \textrm{det}\textbf{J}(D_{\phi}+D_{\psi})^2+(\textrm{tr}\textbf{J})^2D_{\phi}D_{\psi}+2\textrm{tr}\textbf{J}(D_{\phi}+D_{\psi})(J_{11}D_{\psi}+J_{22}D_{\phi})\\
C_{11}&=& 2\textrm{tr}\textbf{J}(D_{\phi}+D_{\psi})\textrm{det}\textbf{J}+(\textrm{tr}\textbf{J})^2(J_{11}D_{\psi}+J_{22}D_{\phi})\nonumber\\
C_{10}&=& \textrm{det}\textbf{J} \left[\textrm{tr}\textbf{J}\right]^2,\nonumber
\end{eqnarray}
and
\begin{eqnarray}
\label{C2}
C_{22}&=& D_{\phi}D_{\psi}\left(D_{\phi}-D_{\psi}\right)^2\nonumber\\
C_{21}&=& \left(J_{11}D_{\psi}+J_{22}D_{\phi}\right) \left(D_{\phi}-D_{\psi}\right)^2 \\
C_{20}&=& J_{11}J_{22}\left(D_{\phi}-D_{\psi}\right)^2.\nonumber
\end{eqnarray}

\subsection*{Details of the Brusselator model}

The Brusselator model involves molecules of two chemical species, whose concentrations 
are here labelled $\phi$ and $\psi$ to make contact with the notation employed in the preceding sections. The species populate the nodes of a network, and therein interact according to the prescribed dynamics, while being allowed to diffuse between adjacent nodes. More specifically, the system evolves according to the general dynamical equations (\ref{eq:reac_dif}) with $f(\phi_i,\psi_i)=1-(b+1)\phi_i+ c\phi_i^2 \psi_i $ and $g(\phi_i,\psi_i)=b \phi_i-c \phi_i^2 \psi _i$, 
$b$ and $c$ acting as external, positive definite, parameters. The Brusselator model admits a unique homogeneous fixed point $\phi^*=1, \psi^*=b/c$, which is stable for $b<1+c$. We additionally require $b>1$, such that  $J_{11}=-1+b>0$, while $J_{22}=-c<0$. The system can develop Turing-like patterns for specific choices of the control parameters ($b,c$). In this investigation we assign the parameters values so to fall outside the domain of instability. The fixed point $(\phi^*, \psi^*)$ is therefore stable versus non homogeneous perturbations, when placed on top of a symmetric graph. 

\subsection*{Network generation}

The WS method is a random graphs generation model that produces graphs with the small world property. In short, given the desired number of nodes $\Omega$ and the mean degree $K$, we first construct a $K$-regular ring lattice, by connecting each node to its $K$ neighbours, on one side only. Then, for every node $i$, we take all edges and rewire them with a given probability $p\in [0,1]$. Rewiring is carried out by replacing the target node, with one of the other nodes, say $j$, selected with a uniform probability, and avoiding self-loops ($i \ne j$ ) and link duplication. Therefore, the mean number of rewired links is $\Omega Kp$. The rewiring is directed: we do not impose a symmetric edge that goes from the selected target node $j$, back to the starting site $i$. Small world networks are found for intermediate values of $p$. Link duplication is avoided here, so that the effects of the directed network (which we wish to demonstrate) are not confused with effects of a weighted network. 

We also implement a variation of this strategy, the so-called Newman-Watts (NW) algorithm. 
In this case, all of the original links in the directed lattice are preserved, and extra links are added. As before, we start from a substrate $K$-regular ring made of $\Omega$ nodes. To make connection with the WS route, the algorithm is designed to add, on average, $\Omega Kp$ long-range directed links, in addition to the links due to the regular lattice \cite{newman1,newman2}. Again $p \in [0,1]$ is a probability to be chosen by the user. The NW algorithm  can be modified so to result in a balanced network (identical number of incoming and outgoing links, per node). To this end, the inclusion of a long range link starting from node $i$ is followed by the insertion of a fixed number ($3$ is our arbitrary choice) of additional links to form a loop which closes on $i$.



\appendix

\section{Topology induced patterns in FitzHugh-Nagumo model}
\label{S1}

As mentioned previously in the main text of this paper, pattern formation in directed networks could be an important step forward in the understanding of the complex behaviour in the neuroscience realm. The underlying spatial support in this scenario is almost always represented by a neural network. Recently, substantial progress has been made in acquiring the architectural structure of the cerebral cortical areas in mammalian brains \cite{sporns}. In all cases the topology of the neural networks studied reflects strong similarity to a directed small-world one \cite{sporns, yu}. So, it is quite reasonable to investigate neural dynamics in the new paradigm of this kind of asymmetric network. For this purpose we will choose a typical mathematical model, to show how patterns, particularly travelling waves, can spontaneously emerge in directed graphs, even when the classical conditions for the instability on a symmetric support are not satisfied. The most preferred candidate in literature, is the FitzHugh-Nagumo model \cite{fitz1, fitz2, nagumo},

\begin{eqnarray}
\frac{du_i}{dt}&=&u_i-u_i^3-v_i+D_u\sum_{j=1}^\Omega\Delta_{ij}u_j\nonumber\\
\frac{dv_i}{dt}&=&c(u_i-av_i-b)+D_v\sum_{j=1}^\Omega\Delta_{ij}v_j
\end{eqnarray}
which is a very simplified model that mimics to some extent the neurons dynamics. In fact, here $u$ is the membrane potential and $v$ the recovery variable. 
The model parameters are $a=0.5$, $b=0.04$, $c=26$, $D_u=0.2$ and $D_v=15$ with a fixed point $(u^*, v^*)=(0.0795, 0.079)$. For this choice of the parameters the homogeneous fixed point is stable to inhomogenous perturbation, when the system is defined on a spatial symmetric support.   

\begin{figure}
\includegraphics[scale=0.8]{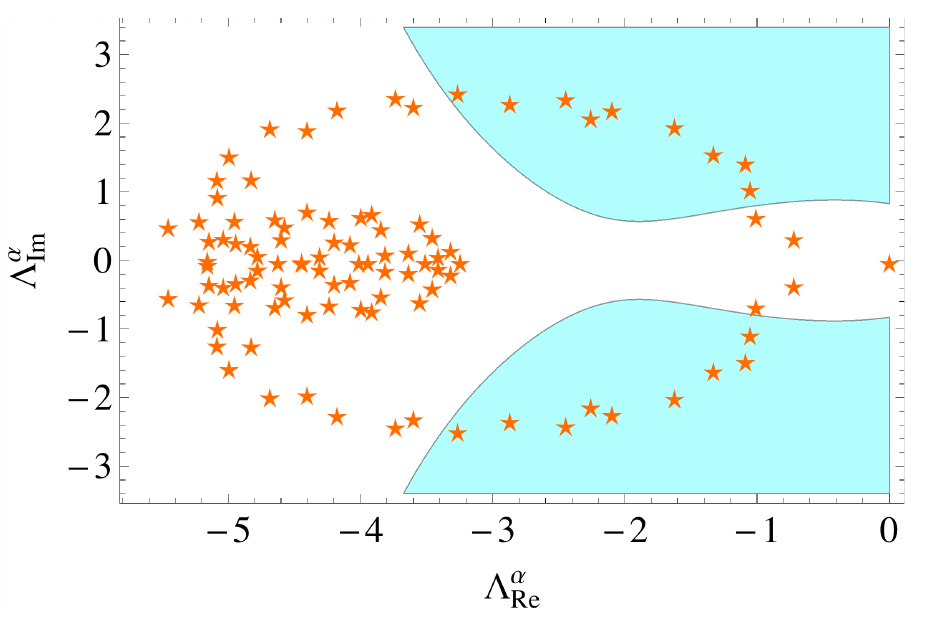}
\caption{Spectral plot of the Laplacians $\Delta_{ij}=W_{ij}-k_i \delta_{ij}$ generated from the Newman-Watts algorithm for $p=0.27$ and network size $N=100$. The shaded area indicates the instability region for the case of the FitzHugh-Nagumo model, where the parameters are $a=0.5$, $b=0.04$, $c=26$, $D_u=0.2$ and $D_v=15$.}
\label{sup_fig1}
\end{figure}

\begin{figure}
\begin{tabular}{cc}
\includegraphics[scale=0.8]{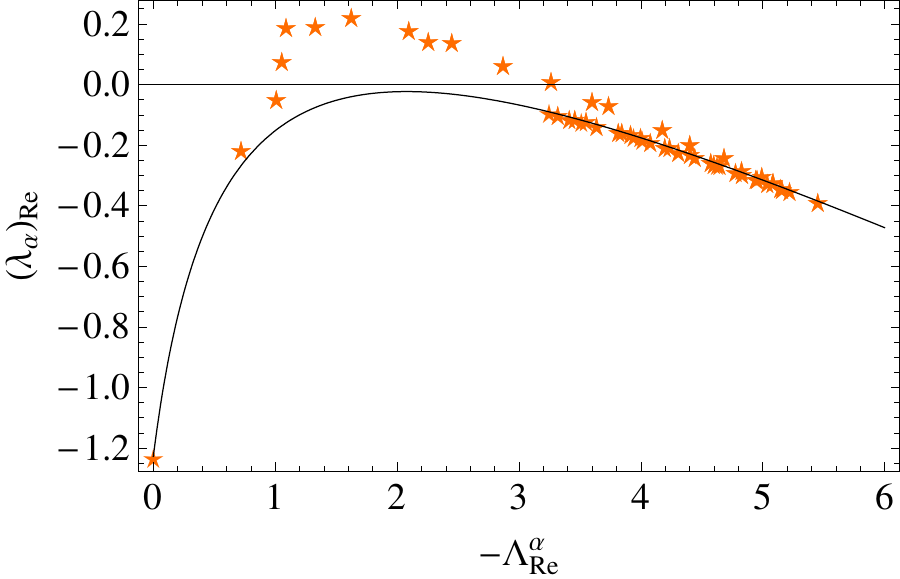}&
\includegraphics[scale=0.8]{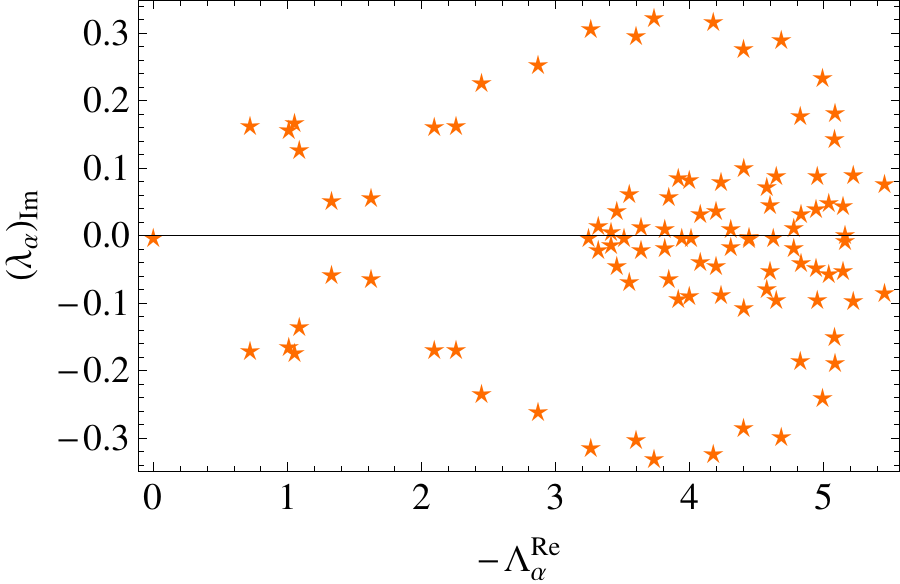}\\
(a) & (b)\\
\end{tabular}
\caption{Panel (a): The real part of the dispersion relation for a Newman-Watts network with a Brusselator dynamic. The network is of size $N=100$, generated with $p=0.27$. Panel (b): The imaginary part of the dispersion relation. The reaction parameter values are given in the caption of Fig.~\ref{sup_fig1}}
\label{sup_fig2}
\end{figure}

\begin{figure}
\includegraphics[scale=0.35]{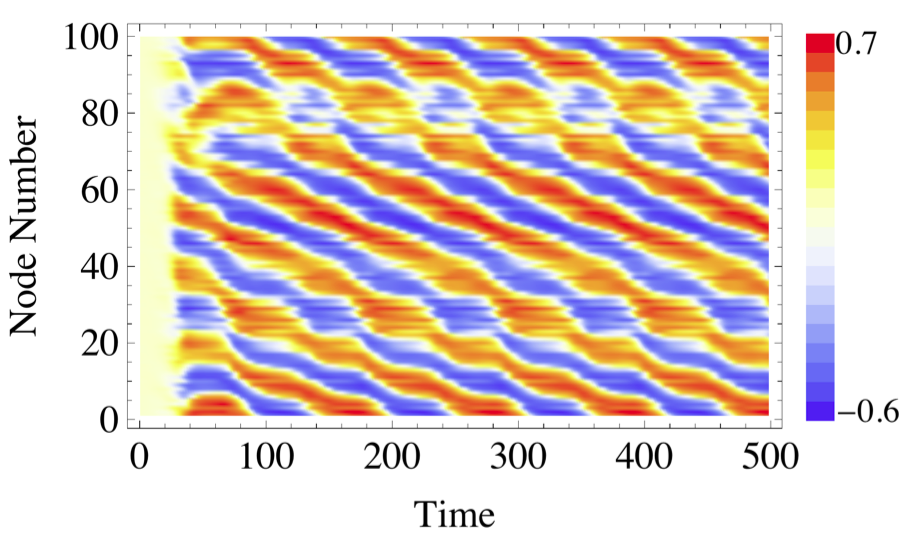}
\caption{Time series for the case of the FitzHugh-Nagumo model on a Newman-Watts network, generated with $p=0.27$. The nodes are ordered as per the original lattice and the reaction parameter values are given in the caption of Fig.~\ref{sup_fig1}}
\label{sup_fig3}
\end{figure}

Below we show the figures corresponding to the FitzHugh-Nagumo model, for both possible choices of the Laplacian operator. Fig.~\ref{sup_fig1} shows the spectrum of the same unbalanced NW small-world with $100$ nodes with $p=0.27$ used in the main text. However, the instability region (shaded domain) has been redrawn for the FitzHugh-Nagumo model. Fig.~\ref{sup_fig2} shows the real and imaginary parts of the dispersion relation, for the same parameter choice. The solid line in the left panel of Fig.~\ref{sup_fig2} stands for the continuous linear instability theory, suggesting that no instability can develop for the model defined on a symmetric support, given the choice of the parameters made. We conclude by presenting the wave pattern for this system in Fig.~\ref{sup_fig3}. Now, the magnitude of the first species can be negative also, as it corresponds to the membrane potential. In Figs. \ref{sup_fig3} 
and \ref{sup_fig4}, the condition of instabilities and the emerging wave are respectively displayed for a balanced NW network, and assuming a purely diffusive transport operator.  

\begin{figure}
\includegraphics[scale=0.8]{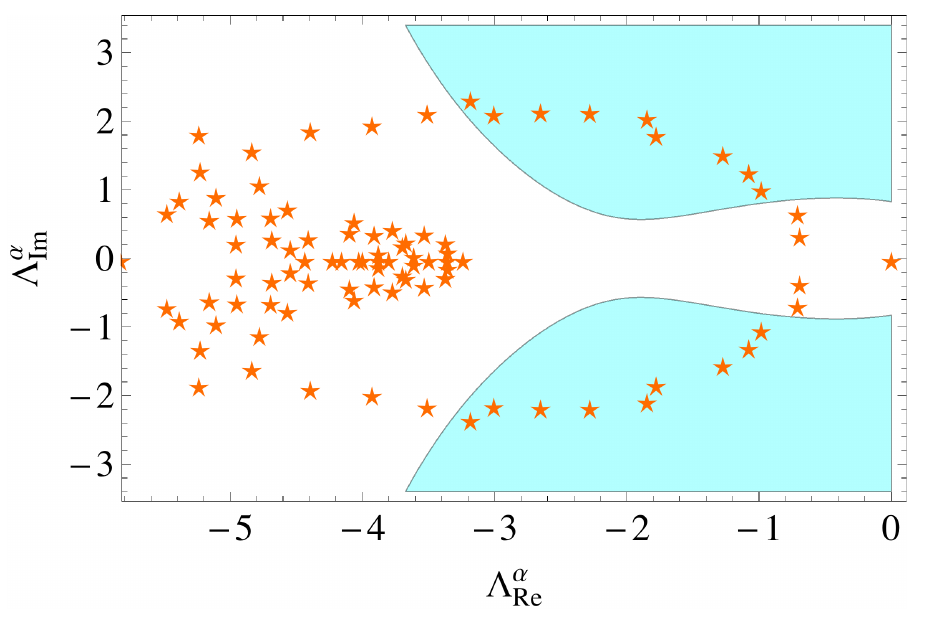}
\caption{Spectral plot of the network Laplacians $\Delta_{ij}=W_{ji}-k_i \delta_{ij}$ generated from the balanced Newman-Watts algorithm for $p=0.27$ and network size $N=100$. The shaded area indicates the instability region for the case of the FitzHugh-Nagumo model, where the parameters are $a=0.5$, $b=0.04$, $c=26$, $D_u=0.2$ and $D_v=15$.}
\label{sup_fig1new}
\end{figure}

\begin{figure}
\begin{tabular}{cc}
\includegraphics[scale=0.8]{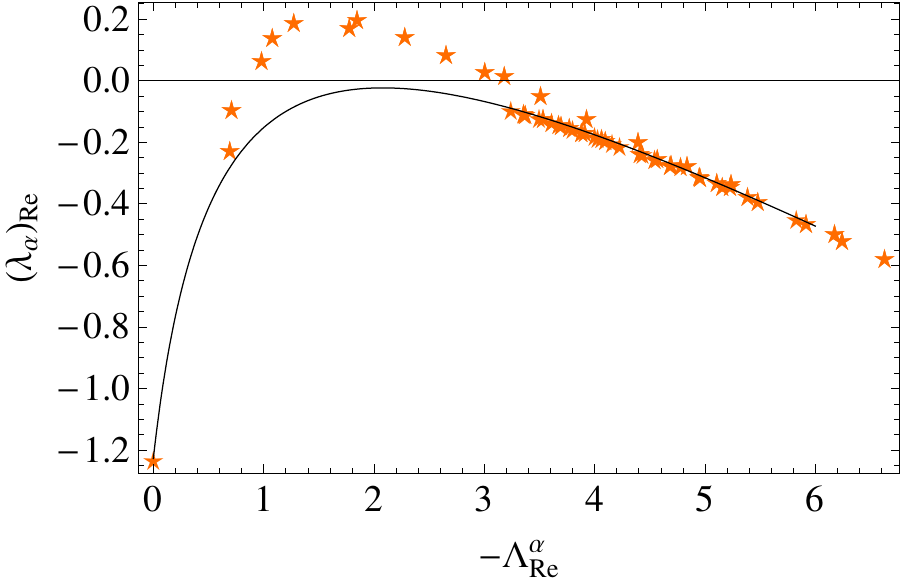}&
\includegraphics[scale=0.8]{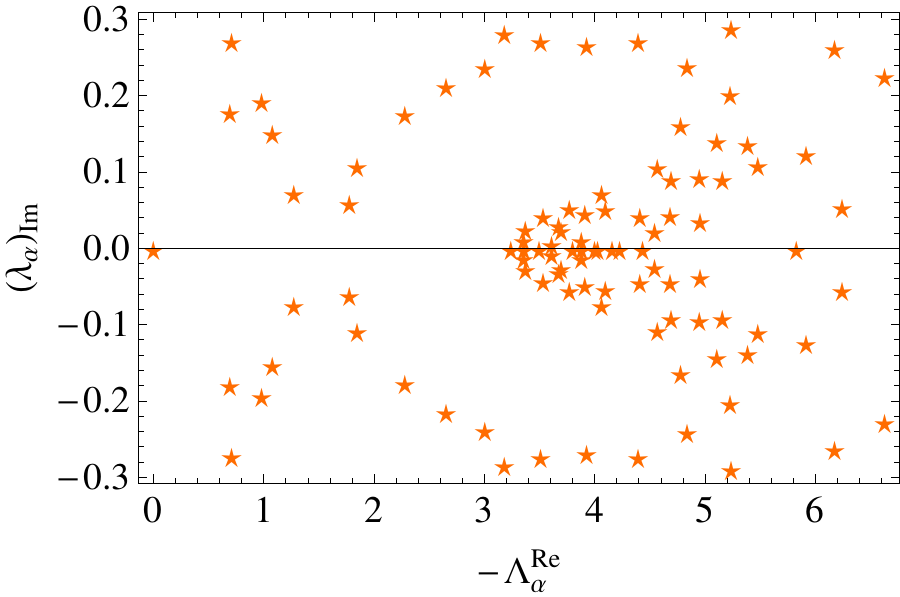}\\
(a) & (b)\\
\end{tabular}
\caption{Panel (a): The real part of the dispersion relation for a  balanced Newman-Watts network with a Brusselator dynamic. The network is of size $N=100$, generated with $p=0.27$. Here, $\Delta_{ij}=W_{ji}-k_i \delta_{ij}$. Panel (b): The imaginary part of the dispersion relation. The reaction parameter values are given in the caption of Fig.~\ref{sup_fig1new}}
\label{sup_fig2new}
\end{figure}

\begin{figure}
\includegraphics[scale=0.35]{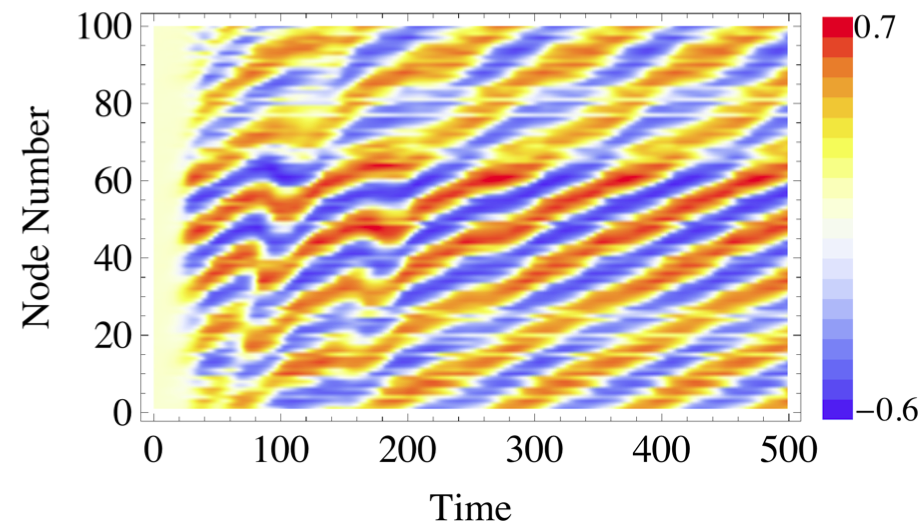}
\caption{Time series for the case of the FitzHugh-Nagumo model on a balanced Newman-Watts network, generated with $p=0.27$. The nodes are ordered as per the original lattice and the reaction parameter values are given in the caption of Fig.~\ref{sup_fig1new}}
\label{sup_fig3new}
\end{figure}

\section{Evolution of wave patterns in small-world networks}  
\label{S2}

The first video file movie1.avi shows the evolution of the pattern for the Brusselator model on a NW small-world with $100$ nodes with $p=0.27$ (see the main text) starting from the homogenous fixed point $\phi^*=1$. For the model parameters refer to the caption of Figure 1 of the main text. Video file movie2.avi shows the travelling wave on the same network, but for the FitzHugh-Nagumo model. For the parameters refer to the previous section. In both cases the vertical axis represents the magnitude of species $\phi$ and the horizontal axis the corresponding node.


\begin{thebibliography}{1}
\bibitem{turing} A. M. Turing, Phil. Trans. R. Soc. Lond. B \textbf{237}, 37 (1952).
\bibitem{mimura} M. Mimura and J. D. Murray, J. Theor. Biol. \textbf{75}, 249 (1978).
\bibitem{maron} J. L. Maron and S. Harrison, Science \textbf{278}, 1619 (1997).
\bibitem{baurmann} M. Baurmann, T. Gross and U. Feudel, J. Theor. Biol. \textbf{245}, 220 (2007).
\bibitem{riet} M. Rietkerk and J. van de Koppel, Trends Ecol. Evol. \textbf{23}, 169 (2008).
\bibitem{meinhardt} H. Meinhardt and A. Gierer, BioEssays \textbf{22}, 753 (2000).
\bibitem{harris} M. P. Harris, S. Williamson, J. F. Fallon, H. Meinhardt and R. O. Prum, Proc. Natl Acad. Sci. USA \textbf{102}, 11734 (2005).
\bibitem{maini} P. K. Maini, R. E. Baker and C-M. Chuong, Science \textbf{314}, 1397 (2006).
\bibitem{bhat} S. A. Newman and R. Bhat, Birth Defects Res. (Part C) \textbf{81}, 305 (2007).
\bibitem{miura} T. Miura and K. Shiota, Dev. Dyn. \textbf{217}, 241 (2000).
\bibitem{murray} J.D. Murray, Mathematical Biology, Second Edition, Springer (1991)
\bibitem{neuron0} J. Wyller, P. Blomquist and G.T. Einevoll, Physica D \textbf{225} 75-93 (2007).
\bibitem{zhab} A. M. Zhabotinsky, M. Dolnik and I. R. Epstein, J. Chem. Phys. \textbf{103}, 10306 (1995).
\bibitem{nakao} H. Nakao and A. S. Mikhailov, Nature Physics \textbf{6}, 544 (2010).
\bibitem{asllani} M. Asllani, T. Biancalani, D. Fanelli, A. McKane, Europ. Phys. J. B \textbf{86}, 476 (2013).
\bibitem{john} W. John, M. Dusi and K. Claffy, Tech. rep., ACM 1st International Workshop on TRaffic Analysis and Classification (TRAC) (2010).
\bibitem{neuron2} E. R. Kandel, J. H. Schwartz and T. M. Jessell. Principles of neural science. Fourth Edition. McGraw-Hill (2000).
\bibitem{neuron3} J. W. Lichtman and W. Denk, Science \textbf{334}, 618 (2011).
\bibitem{neuron4} O. Sporns, G. Tononi and R. K\"{o}tter, PloS Computational Biology \textit{1}, e42 (2005).
\bibitem{connectome} http://www.humanconnectomeproject.org/
\bibitem{biancalani} T. Biancalani, T. Galla and A. J. McKane, Phys. Rev E \textbf{84}, 026201 (2011).
\bibitem{SaberMurray} R. Olfati-Saber and R.M. Murray IEEE Transactions on Automatic Control \textbf{49}(9) 1520-1533 (2004). 
\bibitem{jost} A. Banerjee and J. Jost, Linear algebra and its applications, \textbf{428} (11-12) 3015-3022 (2008).
\bibitem{nonhom} C.N. Angstmann, I.C. Donnelly, B.I. Henry, Phys. Rev E \textbf{87}, 032804 (2013).
\bibitem{ridolfi} L. Ridolfi, C. Camporeale, P. D'Odorico and F. Laio, Europhysics Letters, \textbf{95}, 18003 (2011).
\bibitem{ger} Bell, H. E,  Amer. Math. Monthly \textbf{72}, 292-295, (1965). 
\bibitem{watts} D. J. Watts and S. H. Strogatz, Nature \textbf{393}, 440-442, (1998).
\bibitem{newman1} M. E. J. Newman and D. J. Watts, Phys. Rev. E \textbf{60}, 7332-7342 (1999).
\bibitem{newman2} M. E. J. Newman, SIAM Review \textbf{45}, 167-256 (2003).
\bibitem{sporns} O. Sporns , D. R. Chialvo, M. Kaiser and C. C. Hilgetag, Trends Cogn. Sci. \textbf{8}(9): 418--425, (2004).
\bibitem{yu} S. Yu, D. Huang, W. Singer and D. Nikoli{\'c}, Cerebral Cortex 18 (12): 2891--2901, (2008).
\bibitem{fitz1} R. FitzHugh, Bull. Math. Biophysics, \textbf{17}, 257--278 B (1955).
\bibitem{fitz2} R. FitzHugh, Biophysical J. \textbf{1}, 445-466 (1961).
\bibitem{nagumo} J. Nagumo, S. Arimoto and S. Yoshizawa, Proc IRE. \textbf{278}, 2061--2070, (1962).
 
 \end{thebibliography}
\end{document}